# Superdirective Mixed-Multipole-based Unidirectional Spherical Dielectric Lens Antennas

Asbjørn T. Birch, Samel Arslanagić, and Richard W. Ziolkowski

*Abstract*—Studies of superdirective dielectric lens antennas are reported emphasizing unidirectional properties. The developed antennas are based on the higher-order transverse electric and magnetic modes present in a multilayered sphere realized with high permittivity dielectrics. The superdirective properties of the lenses are empowered by exciting them with a basic unidirectional mixed-multipole antenna. Genetic algorithm (GA) optimization was employed to identify configurations that yielded both large directivity and front-to-back ratio outcomes. The analytical evaluations of several example systems of spheres having different outer radii and numbers of layers are described. Their radiated field patterns are presented along with comparisons with known directivity bounds to confirm their superdirective performance characteristics. Subsequent numerical assessments with commercial full-wave finite element simulations of more realistic versions of several selected examples illustrate their potential for experimental realization.

## I. Introduction

Advanced higher frequency, highly directive antenna systems will enable the realization of a wide variety of current and future wireless communication, sensor, and power transfer applications for the benefit of humanity. They provide a practical means to overcome atmospheric propagation losses associated with those frequencies while facilitating secure channels of information and efficient energy exchanges. While phased arrays can provide the desired high directivities, they are power hungry systems and require large physical areas. Consequently, high directivity, low-loss, compact antennas remain at the forefront of expediting many wireless applications associated with NextG systems, the Internet-of-Things (IoT), and the Internet-of-Everything (IoE). Superdirective radiating elements and/or arrays of them would facilitate these objectives.

The concept of superdirectivity originated in the seminal paper by Oseen [1] in which he suggested a classical (wave-based) model of Einstein's *needle radiation*, i.e., the quantum mechanical photon. It was conceived as a superposition of higher-order multipoles at the same frequency formed analytically as derivatives of the free-space Green's function. This fundamental concept of needle radiation through a solution of Maxwell's equations was demonstrated explicitly in [2].

Asbjørn T. Birch, Samel Arslanagić are with the Department of Space Research and Technology, Technical University of Denmark (DTU), DK-2800, Kgs. Lyngby, Denmark, email: s184742@student.dtu.dk, saar@dtu.dk

Richard W. Ziolkowski is with the Department of Electrical and Computer Engineering, University of Arizona, Tucson, AZ 85721-0104 USA (e-mail: ziolkows@arizona.edu)

There were many theoretical papers published in the last century about superdirectivity and supergain, e.g., [3]–[9]. However, the common consensus then was that superdirective systems were not practical because of major problems associated with them including low radiation resistances; low efficiencies; narrow bandwidths; and sensitivities to design parameters and, hence, fabrication tolerances, as summarized succinctly, e.g., in [7], [8]. A pivotal turning point in superdirectivity was the demonstration in the beginning of this century of practical electrical small (ES) two-element superdirective end-fire monopole arrays, e.g., in [10]–[12]. In the past decade there have been numerous examples of experimentally demonstrated superdirective end-fire arrays of electric elements, e.g., [13]–[20], and magnetic elements, e.g., [21]–[23].

To determine if an antenna system is superdirective or not, the applied electromagnetics (EM) community has relied on developed directivity bounds of radiating and receiving systems in linear, time-invariant (LTI) environments. Uzkov [24] demonstrated that the maximum directivity of a linear end-fire array of $M$ isotropic radiators would be $M^2$ as the distance between the elements approached zero. Harrington used spherical wave expansions of the EM fields to derive the well-known maximum directivity bound $D_{\text{HM}} = N^2 + 2N$ for a system composed of electric and magnetic multipoles whose largest order is $N$ in [25]. Moreover, Harrington also postulated in [25]–[27] that multipoles of order $N > ka$, where $k$ is the free space wavenumber and $a$ is the radius of the smallest sphere enclosing the entire antenna, cannot be effectively used for engineering purposes for a system whose electrically size is $ka$. Consequently, he suggested the size-constrained maximum directivity bound $D_{\text{ka}} = (ka)^2 + 2ka$. This has become a defacto comparison choice for many recent endfire-radiating array articles. Nevertheless, it was recognized by Kildal and Best in [28] that $D_{\text{ka}}$ inappropriately goes to zero as $ka$ does, i.e., the directivity of an ideal electric or magnetic dipole is well-known [29] to be $D_{\text{dipole}} = 1.5 = 1.76$ dB. Consequently, the heuristic bound $D_{\text{KB}} = (ka)^2 + 3$ was suggested because an ideal Huygens dipole antenna (HDA) has the maximum directivity $D_{\text{HDA}} = 3 = 4.77$ dB. The appropriateness of this bound was reinforced in [30]. Finally, it is also well-known [29] that the maximum directivity of an aperture antenna/array whose aperture area $A$ is uniformly excited with a time-harmonic signal whose wavelength in free space is $\lambda_0$ is $D_A = 4\pi A/\lambda_0^2$. As emphasized, e.g., in [7], then a broadside-radiating system is superdirective if its directivity exceeds this value. We note that for a circular aperture of radius $a_{\text{circ}}$ one obtains $D_A = 4\pi(\pi a_{\text{circ}}^2)/\lambda^2 = (ka_{\text{circ}})^2$. These bounds are summarized in Table I for easy reference.

TABLE I: **Directivity Upper Bounds**

| Name | Defining Parameter | $D_{\max}$ |
|---|---|---|
| Harrington $D_{\text{HM}}$ | Highest Order $N$ E & M Multipoles | $N^2 + 2N$ |
| Harrington normal $D_{\text{ka}}$ | Electrical size ka | $(ka)^2 + 2ka$ |
| Kildal-Best $D_{\text{KB}}$ | $ka \leq 1$ $ka > 1$ | 3 $(ka)^2 + 3$ |
| Aperture $D_A$ | Uniformly excited $A$ Circle $A = \pi\, a_{\text{circ}}^2$ | $4\pi A/\lambda_0^2$ $(ka_{\text{circ}})^2$ |

Spherical lens antennas, such as the Luneburg and Maxwell fish-eye lenses, have been employed for many years to realize multi-beam and rapid scanning highly directive systems at microwave and millimeter wave frequencies. These lenses ideally transform the spherical field radiated by a point source into a plane wave over their effective circular output apertures. Their spherical symmetry facilities pointing their output beams in a specified direction simply by rotating the location of the source exciting the sphere to have the maximum of the field it radiates pass through the center of the sphere in the desired direction. Multiple beams in prescribed directions can then be produced by exciting the lens with an array of sources on a fixed radius arc whose fields point towards them. As noted in the brief 1995 review [31], R. K. Luneberg proposed the Luneberg lens in 1944 [32], and their practical applications had been limited because of the weight of the lens and the complexities associated with manufacturing them. This changed soon after as it was found that multiple concentric spherical shells composed of dielectric materials could be used effectively to realize the required radial permittivity profiles. Because the performance of spherical lens antennas are considered with respect to their aperture efficiency (AE), i.e., the ratio of the actual peak gain, $G_{\text{peak}}$, compared to the ideal maximum $G_{\max} = D_A$, and their maximum directivity, $D_{\max}$ value, our main criterion for superdirective performance will be whether or not $D_{\max} > D_A$. Nevertheless, we will also demonstrate that our systems yield $D_{\max}$ values larger than both $D_{\text{ka}}$ amd $D_{\text{KB}}$.

Early optimization techniques such as the genetic algorithm (GA) facilitated the original multiple-shell designs of spherical lense antennas, e.g., Luneburg lenses, achieving acceptable gain and sidelobe performance [33], [34]. With renewed interest they were employed in mobile communication, remote sensing, and radio astronomy systems to achieve high directivities. Many advances in their designs have since ensued and continue to be explored for a variety of NextG applications. Furthermore, as realized more recently for nanoantennas, these dielectric based systems offer quite low material losses. Consequently, remembering that gain is overall efficiency times directivity, high AE systems have been developed in multiple frequency ranges ranging from microwave to optical wavelengths. For instance, the six-shell, 8 $\lambda_0$ diameter spherically stratified Luneburg lens fed by an open-ended waveguide reported in [35] attained a $D_{\max} \sim$ 27.3 dB (AE $\sim$ 85%) at 6.0 GHz with similar system results reported at 50 and 77 GHz. The lightweight metamaterial-inspired realization of the spherical Luneburg lens antenna in [36] was 4.6 $\lambda_0$ in diameter at 3.55 GHz and had a measured $G_{\text{peak}}$ = 22 dBi and AE $\sim$ 76% The investigation of Luneburg-style lenses excited by open-ended waveguides and horn antennas reported in [37] described, for example, a waveguide-excited 9.15 $\lambda_0$ diameter lens with a simulated $D_{\max}$ = 28.5 dBi at 18 GHz with an AE $\sim$ 86%. Luneburg lense at much higher frequencies are considered frequently as types of graded index (GRIN) lens. Examples of Luneburg lenses at terahertz (THz) frequencies include, e.g., a 2D lens realized as a waveguide-based artificial dielectric medium in [38] and a 3D lens realized with 3D printing in polystyrene in [39]. A 3D Luneburg lens at optical frequencies was constructed in [40] with dielectric metamaterials formed by femtosecond laser direct writing.

In this paper we present superdirective unidirectional spherical lens antennas comprised of custom-designed multilayered spheres formed with concentric shells of low and high permittivity dielectrics that are excited by a basic mixed-multipole antenna (MMA) [41], i.e., a source that consists of an electric (ED) dipole and an orthogonal magnetic (MD) dipole whose amplitudes are independently controlled and, therefore, is a variation of the many successful practical ES HDAs reported in [42] which employ a balanced combination. Herein, unidirectional performance will mean that the front-to-back ratio (FTBR), i.e., the ratio of the peak directivity in a specific direction to the directivity in the opposite one, is FTBR $\geq$ 10 dB. This work can be viewed as a three-dimensional (3D) generalization of the successful two-dimensional (2D) superdirective efforts in [43]. The latter was based on an infinite magnetic line source exciting positive and negative permittivity material layers and only transverse magnetic (TM) modes were available since the configuration was 2D. A practical realization has been reported in [44]. The 3D configuration with its low and high permittivity layers provides access to both transverse electric (TE) and TM modes which can be excited by either or both the ED and MD fields radiated by the MMA. Moreover, we elect to restrict the permittivity values to be greater than or equal to that of free space. This choice simplifies the material models, i.e., our examples are in the microwave regime and then dispersive effects do not play a role. Furthermore, the 3D nature of the configurations we consider overcome the restrictions of the 2D physics. The MMA has the ability to excite both electric and magnetic higher-order multipoles in the multilayered spheres that occur near the same frequency. The consequent superposition of their radiated fields leads to the superdirective outcome, i.e., a maximum directivity $D_{\max}$ significantly larger than $D_A$, e.g., 4-7 dB greater, and, hence, AE values on the order of 250-500%.

There have been many works emphasizing similar multipole physics, but with the excitation being either a plane wave, e.g., for microwave frequencies [45]–[48] and for optical frequencies as nanoantennas [49]–[55], or a simple electric dipole [56]–[62]. The simultaneous excitation with both ED and MD fields, as well as them being unbalanced, is innovative. In fact, it will be emphasized that the MMA excitation can be tailored to generally lead to a 4–6 dB enhancement in the maximum directivity over a similar scatterer excited only by one of them.

The specific parameters of the designs were obtained with an in-house genetic algorithm (GA) based optimizer with a cost-function emphasizing large peak directivity and FTBR values. The details of the GA approach we employed are described in Sec. II. Several of the optimized multi-layer dielectric lenses excited by the combination of an ideal ED and MD will be presented in Section III to illustrate these concepts and their superdirective outcomes along with their tailored unidirectional properties. The optimized examples are selected to illustrate the impact of the lens radius, the number of layers, and their permittivities, as well as the performance enhancements obtained by employing the unbalanced MMA source. Several lens configurations are presented in Section IV that were also confirmed with numerical simulations when their multilayered spheres were excited by practical balanced and unbalanced MMA sources. Conclusions are summarized in Section VI. Our work has involved analytical studies and their MATLAB evaluations, related in-house Python-based high performance codes, and commercial software (ANSYS HFSS v21, CST Microwave Studio, and COMSOL Multiphyics). The time-factor $\exp(j\omega t)$, with $\omega$ being the angular frequency $\omega = 2\pi f$ and $t$ being the time, is assumed and suppressed throughout this paper.

## II. SPHERICAL LENS CONFIGURATION AND ANALYSIS

The configuration investigated in this work is that of a multi-layered sphere centered at the origin which consists of $N$ concentric dielectric layers excited by a MMA located in the region exterior of the sphere. It is illustrated in Fig. 1. Each layer is characterized by the permittivity $\varepsilon_i$ and exterior radius $r_i$ for $i = 1, 2..., r_{N-1}, r_N$. Furthermore, only simple dielectrics, i.e., linear and time-invariant (LTE), homogeneous, and isotropic, are considered. The sphere is embedded in free space, which has the permittivity $\varepsilon_0$. All regions are assumed to have a permeability equal to that of free space, $\mu_0$. Because of the spherical symmetry of the sphere, the MMA location can be selected for convenience. It is assumed to be centered at $\mathbf{r}_s = -r_s \hat{\mathbf{x}}$, a distance $r_s$ along the -$x$ axis. The observation point is denoted as $\mathbf{r}$.

For analysis and optimization purposes, the MMA consists on an ideal ED oriented along the +$z$-axis with current moment $I_e\ell_e$ and a co-located in-phase orthogonal ideal MD oriented along the -$y$-axis with current moment $I_m\ell_m$. These directions were choices made simply to immediately streamline the discussion. If the ED and MD are a balanced pair, i.e., if their current moments satisfy $I_m\ell_0 = \eta_0 I_e\ell_0 = \eta_0 I_0\ell_0$, where the wave impedance of free space $\eta_0 = \sqrt{\mu_0/\varepsilon_0}$, then an HDA is formed that radiates a cardioid directivity pattern with its maximum along the +$x$-axis [63]. An unbalanced pair will have the current amplitudes $W I_m = \eta_0 I_e$, where $W$ will denote the ratio ED:MD, i.e., the ratio of the amplitudes of the electric and magnetic currents relative to those of the HDA version. Consequently, $W$ also represents the relative weights of the electric and magnetic dipole contributions to the overall field radiated by the MMA. Therefore, whether the pair is completely balanced or not, the fields radiated by the MMA illuminate the sphere with the aim to have the maximum directivity of this spherical lens system to be along the +$x$-axis.

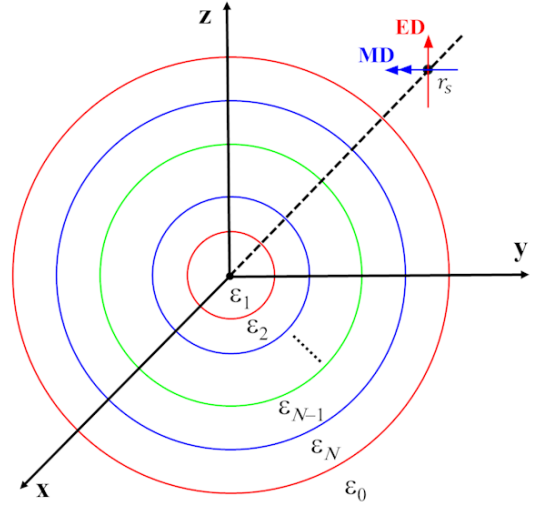

Fig. 1: Spherical lens antenna configuration which consists of an $N$-layered dielectric sphere centered at the origin and excited by a basic MMA located at a distance $r_s$ along the -$x$-axis.

The current densities of the ideal dipoles are

$$\mathbf{J}_{\text{ED}}(\mathbf{r}) = J_{\text{ED}}\,\hat{\mathbf{z}}\,\delta^3(\mathbf{r} - \mathbf{r}_s)$$
$$\mathbf{J}_{\text{MD}}(\mathbf{r}) = -J_{\text{MD}}\,\hat{\mathbf{y}}\,\delta^3(\mathbf{r} - \mathbf{r}_s)$$
(1)

where the current moments $J_{\text{ED}} = I_0\ell_0$ and $J_{\text{MD}} = W\,\eta_0\,I_0\ell_0$. With the noted time convention, the time-harmonic fields radiated by the ED have the well-known Green's function form ( [64], sec. 9.2):

$$\mathbf{E}_{\text{ED}}(\mathbf{r}) = -jk_0\eta_0\,J_{\text{ED}}\left\{\left[\hat{\mathbf{z}} - \left((\hat{\mathbf{z}}\cdot\hat{\mathbf{R}})\,\hat{\mathbf{R}}\right)\right]\right.$$
$$\left. - \left[\hat{\mathbf{z}} - 3\left(\hat{\mathbf{z}}\cdot\hat{\mathbf{R}}\right)\hat{\mathbf{R}}\right]\left(\frac{j}{kR} + \frac{1}{(kR)^2}\right)\right\}G(R)$$
(2)

$$\mathbf{H}_{\text{ED}}(\mathbf{r}) = jk\,J_{\text{ED}}\left[\hat{\mathbf{z}}\times\hat{\mathbf{R}}\right]\left(1 - \frac{j}{kR}\right)G(R)$$

where $R = |\mathbf{r} - \mathbf{r}_s|$ such that $\mathbf{R} = R\hat{\mathbf{R}}$ and the Green's function of the three-dimensional Helmholtz equation in free space is

$$G(R) = \frac{e^{-jk_0 R}}{4\pi R}$$
(3)

and the free-space wave number $k_0 = \omega\sqrt{\varepsilon_0\,\mu_0}$. The corresponding MD fields follow by duality ( [65], sec. 7.2):

$$\mathbf{E}_{\text{MD}} = -jk_0\,J_{\text{MD}}\left[-\hat{\mathbf{y}}\times\hat{\mathbf{R}}\right]\left(1 - \frac{j}{kR}\right)G(R)$$
(4)

$$\mathbf{H}_{\text{MD}} = \frac{-jk_0}{\eta}\,J_{\text{MD}}\left\{-\left(\hat{\mathbf{y}} - (\hat{\mathbf{y}}\cdot\hat{\mathbf{R}})\hat{\mathbf{R}}\right)\right.$$
$$\left. + \left(\hat{\mathbf{y}} - 3(\hat{\mathbf{y}}\cdot\hat{\mathbf{R}})\,\hat{\mathbf{R}}\right)\left(\frac{j}{kR} + \frac{1}{(kR)^2}\right)\right\}G(R)$$

These expressions are the common ones found in textbooks and simply used herein to check the employed vector spherical wave expansion (VSWE) representations of the ED and MD fields.

## A. Vector Spherical Wave Expansion Analysis

The UMMA-multilayered sphere system was first analyzed with the well-known vector spherical wave expansion (VSWE), e.g., [64]–[68]. It is very advantageous for modeling the scattering of incident fields, such as those radiated by the UMMA, from a spherical object such as the $N$-layered dielectric sphere. We employ the VSWE formulation from [57], [69], [70] which facilitated the optimization code development. An arbitrary field is represented as:

$$\mathbf{E}(\mathbf{r}) = \sum_{n=1}^{\infty} \sum_{m=-n}^{n} \frac{1}{j\omega\varepsilon} A_{nm} \mathbf{N}_{nm}^{(c)}(\mathbf{r}) - B_{nm} \mathbf{M}_{nm}^{(c)}(\mathbf{r}) \quad (5)$$

$$\mathbf{H}(\mathbf{r}) = \sum_{n=1}^{\infty} \sum_{m=-n}^{n} \frac{1}{j\omega\mu} B_{nm} \mathbf{N}_{nm}^{(c)}(\mathbf{r}) + A_{nm} \mathbf{M}_{nm}^{(c)}(\mathbf{r})$$

where the coefficients $A_{nm}$ and $B_{nm}$ are determined based on the source of these fields and boundary conditions associated with the region in which they exist. The VSW functions (VSWFs) are:

$$\mathbf{M}_{nm}^{(c)}(\mathbf{r}) = w_n^{(c)}(kr) \left[ \hat{\boldsymbol{\theta}} \, \frac{jm}{\sin\theta} P_n^{|m|}(\cos\theta) - \hat{\boldsymbol{\phi}} \, d_\theta P_n^{|m|}(\cos\theta) \right] e^{jm\phi} \quad (6)$$

$$\mathbf{N}_{nm}^{(c)}(\mathbf{r}) = \hat{\mathbf{r}} \, \frac{n(n+1)}{r} w_n^{(c)}(kr) P_n^{|m|}(\cos\theta) e^{jm\phi} + \frac{1}{r} d_r \left\{ r w_n^{(c)}(kr) \right\} \left[ \hat{\boldsymbol{\theta}} \, d_\theta P_n^{|m|}(\cos\theta) + \hat{\boldsymbol{\phi}} \, \frac{jm}{\sin\theta} P_n^{|m|}(\cos\theta) \right] e^{jm\phi}$$

The superscript $(c)$ identifies the radial spherical wave function appropriate for a given region. In the core of the sphere, which contains the origin, then it is the spherical Bessel function of order $n$, i.e, $w_n^{(1)}(\cdot) = j_n(\cdot)$. If the field is within a layer (bounded region) not containing the origin, then the appropriate VSWFs are a combination of the spherical Bessel functions $w_n^{(1)}(\cdot)$ and spherical Neumann functions $w_n^{(2)}(\cdot) = y_n(\cdot)$ of order $n$ with independent coefficients for each. In the region exterior to the sphere, the fields are outgoing and must decay at infinity. Thus, the appropriate radial wave function for the assumed time-dependence is $w_n^{(3)}(\cdot) = h_n^{(2)}(\cdot)$ corresponding to a spherical Hankel function of the second kind. The functions $P_n^{|m|}(\cos\theta)$ are the associated Legendre functions of the first kind of order $n$ and degree $|m|$. The operators $d_r, d_\theta$ are simplified representations of the differential operators with respect to the subscripted coordinate.

## B. VSWEs of the ED and MD Fields

The fields radiated by the ED and MD located at $\mathbf{r}_s$ have the VSWEs:

$$\mathbf{E}_{\text{ED}}(\mathbf{r}) = \sum_{n=1}^{\infty} \sum_{m=-n}^{n} \frac{1}{j\omega\varepsilon} a_{nm}^{(c)} \mathbf{N}_{nm}^{(c)}(\mathbf{r}) - b_{nm}^{(c)} \mathbf{M}_{nm}^{(c)}(\mathbf{r}) \quad (7)$$

$$\mathbf{H}_{\text{ED}}(\mathbf{r}) = \sum_{n=1}^{\infty} \sum_{m=-n}^{n} \frac{1}{j\omega\mu} b_{nm}^{(c)} \mathbf{N}_{nm}^{(c)}(\mathbf{r}) + a_{nm}^{(c)} \mathbf{M}_{nm}^{(c)}(\mathbf{r})$$

where the expansion coefficients [70]

$$a_{nm}^{(c)} = -jk \Lambda_{nm} \left[ \vec{J}_{\text{ED}} \cdot \mathbf{N}_{n,-m}^{(4-c)}(\mathbf{r}_e) \right] \quad (8)$$

$$b_{nm}^{(c)} = \frac{k^3}{\omega\varepsilon} \Lambda_{nm} \left[ \vec{J}_{\text{ED}} \cdot \mathbf{M}_{n,-m}^{(4-c)}(\mathbf{r}_e) \right]$$

where

$$\Lambda_{nm} = \frac{1}{4\pi} \frac{2n+1}{n(n+1)} \frac{(n-|m|)!}{(n+|m|)!} \quad (9)$$

When $r < r_s$, $c = 1$, while for $r > r_s$, $c = 3$. As can be seen, the dependence of the fields on the source coordinates are contained in these coefficients. The VSWEs of the MD fields are obtained readily using duality and are given by:

$$\mathbf{E}_{\text{MD}}(\mathbf{r}) = -\sum_{n=1}^{\infty} \sum_{m=-n}^{n} \frac{1}{j\omega\varepsilon} d_{nm}^{(c)} \mathbf{N}_{nm}^{(c)}(\mathbf{r}) + c_{nm}^{(c)} \mathbf{M}_{nm}^{(c)}(\mathbf{r}) \quad (10)$$

$$\mathbf{H}_{\text{MD}}(\mathbf{r}) = \sum_{n=1}^{\infty} \sum_{m=-n}^{n} \frac{1}{j\omega\mu} c_{nm}^{(c)} \mathbf{N}_{nm}^{(c)}(\mathbf{r}) - d_{nm}^{(c)} \mathbf{M}_{nm}^{(c)}(\mathbf{r})$$

where the coefficients are now:

$$c_{nm}^{(c)} = -jk \Lambda_{nm} \left[ \vec{J}_{\text{MD}} \cdot \mathbf{N}_{n,-m}^{(4-c)}(\mathbf{r}_m) \right] \quad (11)$$

$$d_{nm}^{(c)} = \frac{k^3}{\omega\mu} \Lambda_{nm} \left[ \vec{J}_{\text{MD}} \cdot \mathbf{M}_{n,-m}^{(4-c)}(\mathbf{r}_m) \right]$$

The combined fields radiated by the ED and MD are then expressed as:

$$\mathbf{E}_{\text{combined}}(\mathbf{r}) = \sum_{n=1}^{\infty} \sum_{m=-n}^{n} \frac{1}{j\omega\varepsilon} (a_{nm}^{(c)} - d_{nm}^{(c)}) \mathbf{N}_{nm}^{(c)}(\mathbf{r}) - (b_{nm}^{(c)} + c_{nm}^{(c)}) \mathbf{M}_{nm}^{(c)}(\mathbf{r}) \quad (12)$$

$$\mathbf{H}_{\text{combined}}(\mathbf{r}) = \sum_{n=1}^{\infty} \sum_{m=-n}^{n} \frac{1}{j\omega\mu} (b_{nm}^{(c)} + c_{nm}^{(c)}) \mathbf{N}_{nm}^{(c)}(\mathbf{r}) + (a_{nm}^{(c)} - d_{nm}^{(c)}) \mathbf{M}_{nm}^{(c)}(\mathbf{r})$$

We note that the transverse magnetic (TM) modes of these expressions are associated with the coefficients $a_{nm}^{(c)}, d_{nm}^{(c)}$, while the the transverse electric (TE) modes are associated with the coefficients $b_{nm}^{(c)}, c_{nm}^{(c)}$.

## C. Far-Field VSWEs

To establish the directivities of the ED and MD source fields and the fields scattered by the sphere, the far-field versions of the VSWEs are necessary. Since all these structures are near the origin and lie in free space, we only need to consider the behavior of the spherical Hankel function of the second kind in the limit $r \to \infty$. In particular, we know, e.g., from [65] that

$$\lim_{r \to \infty} \left[ w_n^{(3)}(kr) = h_n^{(2)}(kr) \right] = j^{(n+1)} \frac{e^{-jkr}}{kr} \quad (13)$$

The associated derivatives follow naturally as

$$\lim_{r \to \infty} \left\{ \frac{1}{r} d_r \left[ r h_n^{(2)}(kr) \right] \right\} = \frac{1}{r} d_r \left\{ j^{(n+1)} \frac{e^{-jkr}}{k} \right\}$$
$$= j^n \frac{e^{-jkr}}{r} \quad (14)$$

Additionally, since the radial component of the VSWFs $\mathbf{N}_{nm}$ decay as $r^{-2}$ when $r \to \infty$, the far-field VSWFs thus take the form:

$$\mathbf{N}_{nm}^{\text{FF}}(\mathbf{r}) = j^n \frac{e^{-jkr}}{r} \left[ \hat{\boldsymbol{\theta}} \, d_\theta P_n^{|m|}(\cos\theta) + \hat{\boldsymbol{\phi}} \frac{jm}{\sin\theta} P_n^{|m|}(\cos\theta) \right] e^{jm\phi}$$

$$\mathbf{M}_{nm}^{\text{FF}}(\mathbf{r}) = j^{(n+1)} \frac{e^{-jkr}}{kr} \left[ \hat{\boldsymbol{\theta}} \frac{jm}{\sin\theta} P_n^{|m|}(\cos\theta) - \hat{\boldsymbol{\phi}} \, d_\theta P_n^{|m|}(\cos\theta) \right] e^{jm\phi}$$
(15)

where the superscript ($c = 3$) has been replaced by the far-field indicator FF. As expected, with the absence of any radial components, the far-field VSWFs will yield a transverse electromagnetic (TEM) field. In particular, with some algebraic manipulations, the $\theta$- and $\phi$-components of the combined electric field in the far field are explicitly:

$$E_{\theta,\text{combined}}^{\text{FF}} = \frac{1}{\omega\varepsilon} \frac{e^{-jkr}}{r} \sum_{n=1}^{\infty} \sum_{m=-n}^{n} j^{(n-1)} e^{jm\phi}$$
$$\times \left[ \xi_{nm} \, d_\theta P_n^{|m|}(\cos\theta) + \zeta_{nm} \frac{jm}{\sin\theta} P_n^{|m|}(\cos\theta) \right]$$
(16)
$$E_{\phi,\text{combined}}^{\text{FF}} = \frac{1}{\omega\varepsilon} \frac{e^{-jkr}}{r} \sum_{n=1}^{\infty} \sum_{m=-n}^{n} j^{(n-1)} e^{jm\phi}$$
$$\times \left[ \xi_{nm} \frac{jm}{\sin\theta} P_n^{|m|}(\cos\theta) - \zeta_{nm} \, d_\theta P_n^{|m|}(\cos\theta) \right]$$

where the combined far-field coefficients, i.e., the coefficients in the free-space region, are

$$\xi_{nm} = a_{nm}^{(3)} - d_{nm}^{(3)}$$
$$\zeta_{nm} = \frac{b_{nm}^{(3)} + c_{nm}^{(3)}}{\eta} \quad (17)$$

The magnetic field follows straightforwardly as $\mathbf{H}_{\text{combined}}^{\text{FF}}(\mathbf{r}) = \hat{\mathbf{r}} \times \mathbf{E}_{\text{combined}}^{\text{FF}}(\mathbf{r})/\eta_0$.

## D. Scattering from the Sphere

When an $N$-layer sphere is illuminated by an incident field, the fields in each of its layers as well as the field it scatters can also be expressed in terms of the VSWFs. In the core region where $r \leq r_1$, the fields are expressed as

$$\mathbf{E}_1 = \sum_{n=1}^{\infty} \sum_{m=-n}^{n} \frac{1}{j\omega\varepsilon} A_{1,nm} \mathbf{N}_{nm}^{(1)} - B_{1,nm} \mathbf{M}_{nm}^{(1)}$$
$$\mathbf{H}_1 = \sum_{n=1}^{\infty} \sum_{m=-n}^{n} \frac{1}{j\omega\varepsilon} B_{1,nm} \mathbf{N}_{nm}^{(1)} + A_{1,nm} \mathbf{M}_{nm}^{(1)}$$
(18)

On the other hand, the fields in each successive layer where $r_{i-1} < r \leq r_i$ are expressed as

$$\mathbf{E}_\ell = \sum_{n=1}^{\infty} \sum_{m=-n}^{n} \frac{1}{j\omega\varepsilon} \left( A_{\ell,nm}^+ \mathbf{N}_{nm}^{(1)} + A_{\ell,nm}^- \mathbf{N}_{nm}^{(2)} \right)$$
$$- \left( B_{\ell,nm}^+ \mathbf{M}_{nm}^{(1)} + B_{\ell,nm}^- \mathbf{M}_{nm}^{(2)} \right)$$
(19)
$$\mathbf{H}_\ell = \sum_{n=1}^{\infty} \sum_{m=-n}^{n} \frac{1}{j\omega\mu} \left( B_{\ell,nm}^+ \mathbf{N}_{nm}^{(1)} + B_{\ell,nm}^- \mathbf{N}_{nm}^{(2)} \right)$$
$$+ \left( A_{\ell,nm}^+ \mathbf{M}_{nm}^{(1)} + A_{\ell,nm}^- \mathbf{M}_{nm}^{(2)} \right)$$

for $\ell = 2, ..., N$. The "+" and "-" superscripts respectively indicate the outgoing and incoming waves that form a standing wave within the layer. Finally, the scattered fields in the region exterior to the sphere, i.e., for $r > r_N$, are expressed as

$$\mathbf{E}_s = \sum_{n=1}^{\infty} \sum_{m=-n}^{n} \frac{1}{j\omega\varepsilon} A_{s,nm} \mathbf{N}_{nm}^{(3)} - B_{s,nm} \mathbf{M}_{nm}^{(3)}$$
$$\mathbf{H}_s = \sum_{n=1}^{\infty} \sum_{m=-n}^{n} \frac{1}{j\omega\varepsilon} B_{s,nm} \mathbf{N}_{nm}^{(3)} + A_{s,nm} \mathbf{M}_{nm}^{(3)}$$
(20)

The various field coefficients are determined by enforcing the electromagnetic boundary conditions (BCs) at each interface, i.e., the tangential components of the total electric and magnetic fields are continuous across each interface. Moreover, their TE and TM constituents are linearly independent. Furthermore, because of the spherical symmetry of each interface, these BCs reduce to relations between the radial functions and their derivatives on both sides of the interface evaluated at its radius. A $4N \times 4N$ scattering matrix $\mathbf{S}_n$ results and has the form:

$$\underline{\mathbf{S}} = \begin{bmatrix} \underline{\mathbf{A}} & \mathbf{0} \\ \mathbf{0} & \underline{\mathbf{B}} \end{bmatrix} \quad (21)$$

where the $2N \times 2N$ matrix $\underline{\mathbf{A}}$ corresponds to the BC-based relations associated with the TM expansion coefficients associated with the multipoles of order $n$:

$$\mathbf{A}_{nm} = \begin{bmatrix} A_{1,nm} \\ A_{2,nm}^+ \\ A_{2,nm}^- \\ \vdots \\ A_{N,nm}^+ \\ A_{N,nm}^- \\ A_{s,nm} \end{bmatrix} \quad (22)$$

and the $2N \times 2N$ matrix $\underline{\mathbf{B}}$ corresponds to the BC-based relations associated with the TE expansion coefficients associated with the multipoles of order $n$:

$$\mathbf{B}_{nm} = \begin{bmatrix} B_{1,nm} \\ B_{2,nm}^+ \\ B_{2,nm}^- \\ \vdots \\ B_{N,nm}^+ \\ B_{N,nm}^- \\ B_{s,nm} \end{bmatrix} \quad (23)$$

Both $\underline{\mathbf{A}}$ and $\underline{\mathbf{B}}$ consist of block matrices formed from the elements of the relations obtained by enforcing the BCs. For the $r = r_1$ and $r = r_N$ interfaces the first two and last two rows of each matrix involve three terms. Each of the other interfaces lead to four terms. For instance, for $r = r_\ell$ for $\ell = 2, ..., N-1$, those matrix elements have the form

$$\begin{bmatrix} j_n(k_i r_i) & d_r|_{r=r_i} [r j_n(k_i r)] \\ y_n(k_i r_i) & d_r|_{r=r_i} [r y_n(k_i r)] \\ -j_n(k_{i+1} r_i) & -\frac{\gamma_i}{\gamma_{i+1}} d_r|_{r=r_i} [r j_n(k_{i+1} r)] \\ -y_n(k_{i+1} r_i) & -\frac{\gamma_i}{\gamma_{i+1}} d_r|_{r=r_i} [r y_n(k_{i+1} r)] \end{bmatrix}^T \quad (24)$$

where for the TM elements, $\gamma_\ell/\gamma_{\ell+1} = [\varepsilon_\ell/\varepsilon_{\ell+1}$ and for the TE elements, $\gamma_\ell/\gamma_{\ell+1} = 1$ because the permeabilities in each layer are the same.

The unknown scattering coefficients than follow from the matrix equation:

$$\mathbf{D} = \underline{\mathbf{S}}^{-1} \, \mathbf{\Omega} \quad (25)$$

where coefficient vector $\mathbf{D} = [\mathbf{A} \, \mathbf{B}]^T$ and the excitation vector $\mathbf{\Omega} = [\mathbf{\Omega}^{TM} \mathbf{\Omega}^{TE}]^T$ follows from the ED and HD fields at $r = r_N$, i.e.,

$$\mathbf{\Omega}^{TM} = \begin{bmatrix} 0 \\ 0 \\ \vdots \\ \left[ a_{nm}^{(1)} - d_{nm}^{(1)} \right] j_n(k_0 r_N) \\ \frac{\varepsilon_N}{\varepsilon_0} \left[ a_{nm}^{(1)} - d_{nm}^{(1)} \right] d_r|_{r=r_N} [r j_n(k_0 r)] \end{bmatrix} \quad (26)$$

and

$$\mathbf{\Omega}^{TE} = \begin{bmatrix} 0 \\ 0 \\ \vdots \\ \left[ b_{nm}^{(1)} + c_{nm}^{(1)} \right] j_n(k_0 r_N) \\ \left[ b_{nm}^{(1)} + c_{nm}^{(1)} \right] d_r|_{r=r_N} [r j_n(k_0 r)] \end{bmatrix} \quad (27)$$

Once the scattered field coefficients are obtained, the total electric far field, including the ED, MD, and scattered field contributions, can be written as:

$$\mathbf{E}_{\text{tot}}^{\text{FF}}(r, \theta, \phi) = \left[ F_{\theta, \text{tot}}(\theta, \phi) \, \hat{\mathbf{\theta}} + F_{\phi, \text{tot}}(\theta, \phi) \, \hat{\mathbf{\phi}} \right] \frac{e^{-jkr}}{r} \quad (28)$$

where the far-field pattern functions:

$$F_{\theta, \text{tot}}(\theta, \phi) = \frac{1}{\omega \varepsilon} \sum_{n=1}^{\infty} \sum_{m=-n}^{n} j^{(n-1)} e^{jm\phi}$$
$$\times \left[ \xi_{nm, \text{tot}} \, d_\theta P_n^{|m|}(\cos\theta) + \zeta_{nm, \text{tot}} \frac{jm}{\sin\theta} P_n^{|m|}(\cos\theta) \right] \quad (29)$$

$$F_{\phi, \text{tot}}(\theta, \phi) = \frac{1}{\omega \varepsilon} \sum_{n=1}^{\infty} \sum_{m=-n}^{n} j^{(n-1)} e^{jm\phi}$$
$$\times \left[ \xi_{nm, \text{tot}} \frac{jm}{\sin\theta} P_n^{|m|}(\cos\theta) - \zeta_{nm, \text{tot}} \, d_\theta P_n^{|m|}(\cos\theta) \right]$$

and the total VSWE coefficients are:

$$\xi_{nm, \text{tot}} = a_{nm}^{(3)} - d_{nm}^{(3)} + A_{s,nm}$$
$$\zeta_{nm, \text{tot}} = \frac{b_{nm}^{(3)} + c_{nm}^{(3)} + B_{s,nm}}{\eta_0} \quad (30)$$

### E. Figures of Merit

The pattern function representation of the fields is very convenient for assessing the far-field performance of the MMA-excited, spherical lens system. The total time-averaged radiated power is typically obtained as [65]

$$P_{\text{tot}} = \oiint_{S_\infty} \mathbf{S} \cdot d\mathbf{s} = \int_{\theta=0}^{\pi} \int_{\phi=0}^{2\pi} \mathbf{S} \cdot \hat{\mathbf{r}} \, r^2 \sin\theta \, d\phi \, d\theta \quad (31)$$

where $S_\infty$ is a closed spherical surface whose radius is well into the far-field of all the sources of the radiated fields. The far-field expression of the Poynting's vector is:

$$\mathbf{S}^{\text{FF}}(\mathbf{r}) = \frac{1}{2} \, \text{Re} \left\{ \mathbf{E}_{\text{tot}}^{\text{FF}} \times \left[ \mathbf{H}_{\text{tot}}^{\text{FF}} \right]^* \right\} = \frac{1}{2\eta} \left| \mathbf{E}_{\text{tot}}^{\text{FF}}(\mathbf{r}) \right|^2 \hat{\mathbf{r}}$$
$$= \frac{1}{2\eta} \frac{1}{r^2} \left( |F_{\theta, \text{tot}}|^2 + |F_{\phi, \text{tot}}|^2 \right) \hat{\mathbf{r}} = (S_\theta + S_\phi) \, \hat{\mathbf{r}} \quad (32)$$

where the asterisk (*) denotes the complex conjugate. Hence,

$$P_{\text{tot}} = \int_{\theta=0}^{\pi} \int_{\phi=0}^{2\pi} (S_\theta + S_\phi) \, r^2 \sin\theta \, d\phi \, d\theta = P_\theta + P_\phi. \quad (33)$$

After both integrations and several manipulations, the total power associated with the $\theta$-component in the VSWF form is:

$$P_\theta = \frac{1}{\eta} \frac{\pi}{(\omega\varepsilon)^2} \sum_{n=1}^{\infty} \sum_{m=-n}^{n} \frac{(n+|m|)!}{(n-|m|)!} \left[ \left( \frac{2n(n+1)}{2n+1} - |m| \right) \right.$$
$$\left. \times |\xi_{nm, \text{tot}}|^2 + |m| \, |\zeta_{nm, \text{tot}}|^2 \right] \quad (34)$$

In the same manner, the power associated with the $\phi$-component in the VSWF form is:

$$P_\phi = \frac{1}{\eta} \frac{\pi}{(\omega\varepsilon)^2} \sum_{n=1}^{\infty} \sum_{m=-n}^{n} \frac{(n+|m|)!}{(n-|m|)!} \left[ |m| \, |\xi_{nm,\text{tot}}|^2 \right.$$
$$\left. + \left( \frac{2n(n+1)}{2n+1} - |m| \right) |\zeta_{nm,\text{tot}}|^2 \right]. \quad (35)$$

Consequently, the total power radiated in VSWF form is:

$$P_{\text{tot}} = \frac{1}{\eta} \frac{\pi}{(\omega\varepsilon)^2} \sum_{n=1}^{\infty} \sum_{m=-n}^{n} \frac{(n+|m|)!}{(n-|m|)!} \frac{2n(n+1)}{2n+1}$$
$$\times \left( |\xi_{nm,\text{tot}}|^2 + |\zeta_{nm,\text{tot}}|^2 \right) \quad (36)$$

Since $[\eta_0 \, (\omega\varepsilon_0)^2] = k_0 \, \omega\varepsilon$ and $\varepsilon_0 \, \eta_0^2 = \mu_0$, this expression can be expanded to emphasize the various contributions to it:

$$P_{\text{tot}} = \frac{\pi}{\omega k} \sum_{n=1}^{\infty} \sum_{m=-n}^{n} \frac{(n+|m|)!}{(n-|m|)!} \frac{2n(n+1)}{2n+1} \quad (37)$$
$$\times \left[ \frac{|a_{nm}^{(3)} - d_{nm}^{(3)} + A_{s,nm}|^2}{\varepsilon_0} + \frac{|b_{nm}^{(3)} + c_{nm}^{(3)} + B_{s,nm}|^2}{\mu_0} \right]$$

The directivity represents the ratio of the radiation intensity into the far-field in a particular direction to the average power intensity in all directions [29]:

$$D(\theta,\phi) = \lim_{r \to \infty} \frac{r^2 \, \hat{\mathbf{r}} \cdot \mathbf{S}^{\text{FF}}}{P_{\text{tot}}/4\pi} = \frac{2\pi}{\eta_0} \frac{\left( |F_{\theta,\text{tot}}|^2 + |F_{\phi,\text{tot}}|^2 \right)}{P_{\text{tot}}} \quad (38)$$

As was indicated, the aim is to maximize the directivity radiated by the UMAA-spherical lens system along the $+x$-axis, i.e., to obtain the largest possible $D_{\max} = D(\theta = \pi/2, \phi = 0)$. Both the $D_{\max}$ and FTBR = $D_{\max}/D(\theta = \pi/2, \phi = \pi)$ will be quoted in decibels (dB), as will the first sidelobe level (SLL), i.e., the difference between $D_{\max}$ and the next largest peak value in the directivity pattern. These figures-of-merit will be considered in all cases reported.

## III. OPTIMIZATION OF MMA-EXCITED MULTI-LAYERED DIELECTRIC SPHERES

The specific parameters of the MMA-excited spherical lens antennas were obtained with an in-house genetic algorithm (GA) based optimizer with a cost-function emphasizing large peak directivity and FTBR values. To this end, the analysis and optimization were performed used Python with relevant packages. The most important packages were: NumPy for fast computations of arrays [71], [72]; SciPy for matrix inversion, special functions, and interpolation [73]; Pymoo for the multi-objective optimization algorithm NSGA-II [74]; and Matplotlib for plotting [75]. The main part of the code revolved around the multi-layered sphere, which was constructed as a class taking in a range of different arguments. When the class is instantiated, all the relevant computations can then take place. This includes the computations of the fields, directivity, radiated power, FTBR and SLL.

### A. Details of the Optimization Approach

An optimization approach was employed to maximize the directivity in the $+x$-direction while also maximizing the FTBR. The parameters considered in the approach included the permittivities and radii of each layer as well as the radial distance $r_s$ to the source and the ED:MD weighting factor $W$. Therefore, the total number of variables is $2N + 2$ where $N$ is the number of layers.

The optimization parameters are represented by the row vector:

$$\mathbf{x} = \begin{bmatrix} \varepsilon_{r1} & \ldots & \varepsilon_{rN} & r_1 & \ldots & r_N & W & r_s \end{bmatrix} \quad (39)$$

The objective functions are represented by the column vector:

$$\mathbf{f}(\mathbf{x}) = \begin{bmatrix} f_1(\mathbf{x}) \\ f_2(\mathbf{x}) \end{bmatrix} = \begin{bmatrix} D(\theta = 90°, \phi = 0°) \\ \text{FTBR} \end{bmatrix} \quad (40)$$

Inequality constraints had to be imposed on the radius of each layer, i.e., $r_i < r_{i+1}$ for $i = 1, 2, ..., N - 1$. Furthermore, upper and lower bounds were imposed on the system parameters, denoted by superscript U and L, respectively. Thus, $\varepsilon_r^{\text{L}} \leq \varepsilon_{ri} \leq \varepsilon_r^{\text{U}}$, $r^{\text{L}} \leq r \leq r^{\text{U}}$, $W^{\text{L}} \leq W \leq W^{\text{U}}$, and $r_s^{\text{L}} \leq r_s \leq r_s^{\text{U}}$.

The consequent multiple objective optimization problem is then be written concisely as:

$$\begin{aligned} \max_{\mathbf{x}} \quad & \mathbf{f}(\mathbf{x}) \\ \text{s.t.} \quad & r_j < r_{j+1}, \qquad j = 1, 2, ..., N - 1, \\ \text{where} \quad & \varepsilon_r^{\text{L}} \leq \varepsilon_{ri} \leq \varepsilon_r^{\text{U}}, \\ & r^{\text{L}} \leq r_i \leq r^{\text{U}}, \qquad i = 1, 2, ..., N, \\ & W^{\text{L}} \leq W \leq W^{\text{U}}, \\ & r_s^{\text{L}} \leq r_s \leq r_s^{\text{U}} \end{aligned} \quad (41)$$

A plethora of optimization algorithms exist. Evolutionary algorithms (EAs) such as particle swarm optimization [76], [77] and genetic algorithms [78] have been utilized in similar electromagnetic problems. More conventional optimization techniques such as convex optimization [43] and minmax optimization [79] have also been employed. Because it has been used effectively to model spherical lens systems such as Luneburg lenses, a GA suitable for multi-objective optimization was selected. Specifically, it is a non-dominated sorting genetic algorithm called NSGA-II [80] implemented in the Python package [74]. One of the main reasons for choosing GA (or any evolutionary algorithm for that matter) is its ability to explore a large parameter-space by invoking randomness. This randomness feature also makes them more suitable for finding the global optimum, although this outcome is not guaranteed. A main drawback of EAs is the number of intensive computations needed to find the optimum. It is much more computationally expensive compared to deterministic algorithms. Nonetheless, they are more versatile and work on almost any type of objective function [81].

*1) Genetic Algorithm Overview:* Genetic algorithms work based on principles adopted from biological evolution in which the strongest and most fit population survives to the next generation [82]. They iteratively apply evolutionary operations: selection, crossover, and mutation, to gradually improve the solution until a satisfactory outcome is uncovered. It can be adapted to be an even better fit within a given application environment [83].

The GA begins by initializing the first population randomly based on the bounds of each optimization variable. A very basic outline of a GA is illustrated in Fig. 2. The initial population is a $N_{\text{pop}} \times N_{\text{var}}$ matrix where $N_{\text{pop}}$ is the size of the population and $N_{\text{var}}$ is the number of optimization variables. Our choices for the population size followed practical experience with those discussed in [83]. Each row of the matrix corresponds to a single individual of the population. After the initialization the objective function is computed for each individual in the population. Next, the cost (i.e., the fitness of the individuals when each objective function – the $D_{\text{max}}$ or the FTBR is considered) – is sorted, and only the most fit individuals make it to the mating pool. From there the fittest individuals are selected to generate offsprings for the next generation, also known as the crossover operation. The offspring generation is based on adding together two individuals with each having been multiplied by some weight. Lastly, before a new evaluation of the objective functions is performed on the new generation, mutations are allowed to occur. They introduce some randomness into the newly created population. The reason for this mutation stage is to avoid the algorithm from focusing on a local minimum/maximum, i.e., to avoid letting the optimizer converge before a true optimal solution is found.

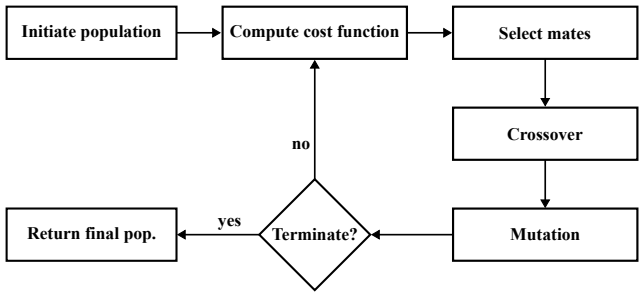

Fig. 2: Simple flowchart describing the working principles of a GA.

*2) The NSGA-II Algorithm:* As previously noted, our optimizations use NSGA-II [80]. This is a mulit-objective GA which is based on non-dominated sorting, i.e., no other individual in the population outperforms a specific individual. Thus, it is not possible to increase the cost of one objective function without decreasing the cost of the other. These specific individuals are the Pareto optimal solutions [82]. A Pareto front is constructed based on this approach to attain all Pareto optimal solutions for a given problem [83].

While other termination/convergence criterions are available, we chose to enforce it to be the maximum number of generations, $N_{\text{gen}}$. The NSGA-II and GAs in general allow for many options during the optimization process including crossover and mutation probability. Nevertheless, the default values from [74] were used in our implementation. Thus, the population size and number of generations are the only parameters that have been changed to obtain the best results. Other possibilities were not explored since this study was focused on the physics associated with the mixed-multipole performance of the UMMA-spherical lens system.

*3) High-Performance Cluster and Parallelization:* Our initial optimization attempts were performed sequentially, i.e., each individual of a population was calculated sequentially on a personal computer. This approach led to extremely large optimization times for a relatively small number of solutions. Therefore, the possibility of using a high-perfromance cluster at the Technical University of Denmark was explored to perform the computations could be done in parallel. It successfully led to a greatly enhanced performance of the optimization algorithm both in computation time and for allowing much larger population sizes and number of generations to be explored. The results obtained were also consistently better in the sense that a larger parameter space was readily explored. Nevertheless, the optimizations still took a long time to complete once an increasing amount of layers was considered.

## IV. SUPERDIRECTIVE MMA-EXCITED SPHERICAL DIELECTRIC LENS ANTENNAS

Our optimization studies emphasized the superdirective and unidirectional properties of several sets of MMA-excited spherical dielectric lens antennas based on the sphere maximum radius $r_{\text{max}} = r_N$ and different numbers of layers for each. We considered $N$ for $2, 3, ..., 10$ layers with $r_{\text{max}} = 8.5, 10.0, 12.0$ mm. Because of experiences with the roll out of previous mobile (wireless) generations and the various practical shortcomings encountered with millimeter-wave systems, telecom companies, e.g., Nokia [84], have begun seriously considering the 7-15 GHz portion of the FR-3 frequency band for a dedicated 6G launch spectrum. Consequently, the target frequency in all cases considered herein was the X-band frequency 10 GHz within the FR-3 band. The optimization studies examined how key parameters such as the population size and the number of generations influenced the ability of the GA-based algorithm to achieve the desired performance characteristics. Several selected configurations are presented that were analyzed in detail to understand the effects of the ED:MD ratio and to assess the impact of rounding the optimized parameters from sixteen to four significant digits. Moreover, several of the GA generated and digit-truncated cases were simulated with a real finite-sized MMA source replacing the ideal ED-MD version in HFSS.

### A. Initial VSWE Numerical Tests with the In-house Python Program

As noted, the numerical implementation of any VSWE representation requires a truncated version of the infinite sum of the order of the multipoles to order $n = M$. This value was determined primarily by comparing the fields radiated by the ED and MD in free space calculated with the VSWE and the Green's function expressions for several source locations away from the origin. The choice of $M$ was refined to avoid extremely large computation times through comparisons of the VSWF results for a MMA-excited multi-layered sphere case with those obtained with the full-wave commercial finite element simulator COMSOL Multiphysics. Note that the relative permittivities of each layer, i.e., the dielectric constant $\varepsilon_r = \varepsilon/\varepsilon_0$, and the ED:MD ratio, $W$, will be given for all cases below. All radii and the source location will be quoted in millimeters.

The first $r_s$ considered was 15.0 mm. With the ED:MD ratio being $W = 1$, the relative error between the VSWE and Green's function (reference case) calculated field components was below 5% for the fields at 8.5 and 10.0 mm, i.e., at the nearest point on the 8.5 and 10 mm spheres, with $M = 15$. On the other hand, it was between 10-20% at 3 mm from the source, i.e., on the nearest point on the 12 mm sphere. We elected to use $M = 15$ in the majority of optimizations of all three cases to be consistent and to save computational times. The order of the higher order TE and TM modes that made a significant contribution to the far-field results were found to be $n = 5$ and below. However, for the extreme $D_{\max}$ (5- and 10-layer) 12.0 mm cases discussed below we opted to use $M = 30$ because the relative error in the VSWE results were then less than 3% at 3 mm from $r_s$. This choice obviated any concerns that modes of order higher than 5 might be responsible for their extreme performance characteristics. Similarly, we opted to also use $M = 30$ for the large physical size case (i.e., $r_N = 1.0\,\lambda = 30$ mm at 10 GHz). With the source distance being $r_s = 35$ mm, the relative error at 5 mm from $r_s$ was then less than 5%. Because it was initially anticipated that more higher order modes might become viable in this physically large case, the choice of $M = 30$ prevented missing their presence and, hence, any of theri contributions to the overall performance.

The corresponding accuracy of the VSWE results in the near field of the lens was also evaluated by comparing their predicted values to a COMSOL Multiphysics simulation of the same 3-layer sphere. The parameters of this comparison were: $\varepsilon_{r1} = 1.0$, $\varepsilon_{r2} = 25.0$, $\varepsilon_{r3} = 4.0$, $r_1 = 2.0$ mm, $r_24 = 6.0$ mm, $r_3 = 8.0$ mm. The MMA was again assumed to be an HDA, i.e., $W = 1.0$. It was located at $\mathbf{r}_s = -12.0\,\hat{\mathbf{x}}$. A PML absorbing boundary condition was employed. The source frequency was 10.0 GHz.

With $M = 15$ the VSWE expressions yielded $D_{\max} = 8.13$ dB and FTBR = 11.82 dB. The COMSOL predicted values were $D_{\max} = 8.09$ dB and FTBR = 12.29 dB. The $D_{\max}$ values were 0.5% different; the FTBR values were 3.98% different. The difference in the FTBR values was larger because of the finite mesh cell size in the COMSOL project which impacted the resolution of the backlobe.

This initial MMA-spherical lens case was determined simply by trial-and-error. It was the initial test case to determine whether or not the MMA source significantly improved the outcomes. In fact, it was found that $D_A = 4.49$ dB, $D_{\max} > D_A$ by 3.64 dB. Moreover, the $AE = 281\%$. It was the first superdirective example.

### B. GA Optimized Systems

Given the importance of $D_A$, it is immediately obvious that the maximum radius of the $N$-layered sphere $r_{\max} = r_N$ has a substantial influence on the maximum achievable directivity of a dielectric spherical lens antenna. The target frequency was 10 GHz, i.e., $\lambda_0 \sim 30$ mm, for the microwave systems. Because the Pareto fronts generated by the optimizer can change significantly when $r_s$ and $W$ are varied, we elected to study three subwavelength diameter spheres, i.e., spheres with their $r_{\max}$ being 8.5, 10.0, and 12.0 mm. The common cutoff order $M = 15$ was applied for these subwavelength $r_{\max}$ cases.

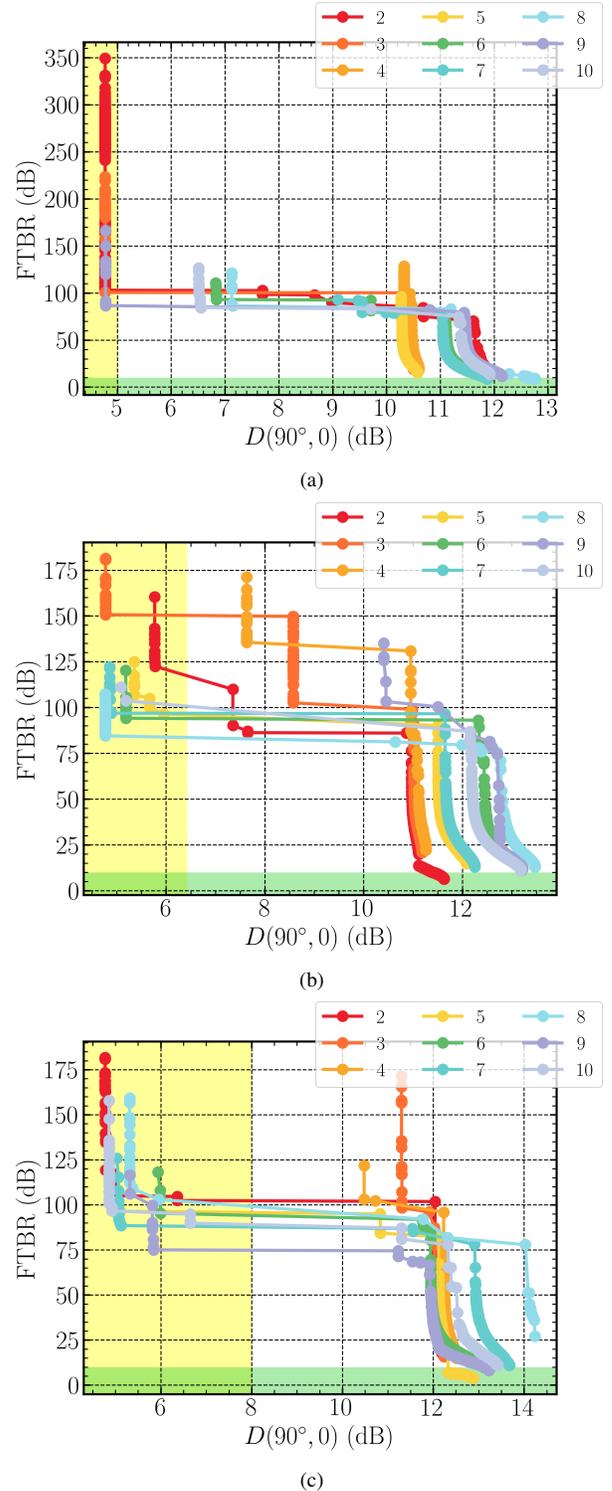

Fig. 3: Pareto fronts obtained by the GA algorithm when the maximum radius of the lens is (a) 8.5, (b) 10.0, and (c) 12.0 mm. The horizontal light green shaded region represents FTBR values between 0 and 10.0 dB. The vertical light yellow shaded region represents $D_{\max}$ values between 0 and $D_A$ for each case. Superdirective unidirectional cases are all those not within either shaded region.

While we have tried optimizations that restricted the maximum dielectric constants to 25, 50, and 100, our initial foray had $\varepsilon_r \in [1, 100]$. Moreover, the ED : MD ratio was bound by $W \in [0, 5]$, and the source location was constrained to $r_s = [r_{\max}, 1.5\, r_{\max}]$. The population and generation numbers were taken to be $N_{\text{pop}} = 1200$ and $N_{\text{gen}} = 500$ based upon discussions in [83].

The Pareto fronts obtained for these three sphere sizes with the number of layers in each case being $N = 2, 3, ..., 10$ are shown in Fig. 3. The region above the horizontal light green shading whose FTBR represent 0 to 10.0 dB indicates the unidirectional solutions. Superdirective cases are those found in the region to the right of the vertical light yellow region which denotes the directivity values from 0 to $D_A$. This right vertical boundary of this shaded region for the 8.5, 10.0, and 12.0 mm cases at 10.0 GHz is $D_A = 5.01$, 6.43, and 8.01 dB, respectively, since $a_{\text{circ}} = r_{\max}$.

There are very clearly a multitude of solutions in which these MMA-spherical lens antennas are simultaneously superdirective and unidirectional. As anticipated, for instance, from [43], [85], more layers are generally seen to lead to higher directivity. Moreover, larger effective output apertures, i.e., larger spheres, also provides access to higher directivities. While the large jumps in the Pareto fronts are strange looking, they are valid. Similar behavior can be seen in the seminal paper on the NSGA-II algorithm [80]. As we realized after these initial results, the parameter space of our problems is huge and the algorithm struggled to find a global optimum even allowing for huge simulation times, especially as the number of layers increased.

Intuitively, having a larger initial population and more generations should yield better results due to the larger number of possible solutions being explored. Nevertheless, a number of times during the optimization, we found worse results followed an increase of $N_{\text{gen}}$ and $N_{\text{pop}}$. They were caused by as associated large increase in the number of local minima/maxima. If the optimizer did not approach a global maximum, it became stuck near an undesirable local maximum. This behavior became an issue because an increase in the number of solutions led to an increase in time with checking whether the solutions were actually desirable or not. Furthermore. while Figs. 3(a)-(c) do indicate the existence of many superdirective solutions, they do not mean that a global maximum was found. Simply, since the initial population is randomly generated within the specified bounds, the optimizer does get stuck near local maxima.

Due to the sheer amount of results obtained per optimization attempt, picking out the best cases based on directivity, FTBR, SLL proved to be difficult. The following are highlighted due to their favorable characteristics in either directivity and FTBR. Cases whose resilience to rounding off the optimized parameters to a practical number of digits are the focus of the HFSS simulation cases discussed in the next section. Note that the optimizer's predicted directivity plots are reported in dB by default and values less than -35 dB are mapped to that value to simplify the presented patterns.

### C. 10-Layer Extreme Directivity Case

The maximum directivity achieved throughout all optimization attempts of the three sphere sizes was attained with a 10-layer 12 mm sphere. The parameters of the GA algorithm were set at $N_{\text{pop}} = 2000$, $N_{\text{gen}} = 1200$. The relative permittivities of the layers were constrained to the interval $\varepsilon_r \in [1, 50]$. The Pareto front of this optimization process is shown in Fig. 4. Notably, recalling that $D_A = 8.01$ dB in this case, a significant number of solutions do achieve superdirective behavior, particularly several having directivities around 15.0 dB. Moreover, many have very large FTBR values, e.g., 100 dB.

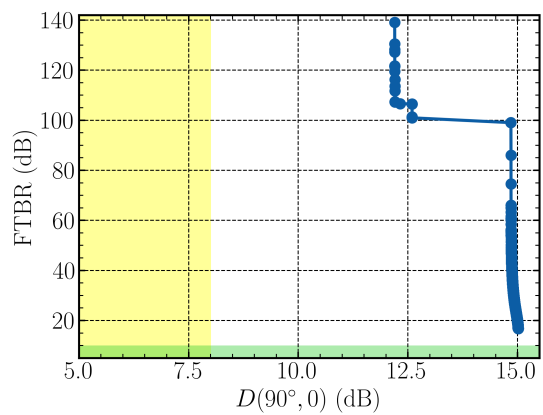

Fig. 4: Pareto front of the maximum directivity of the 10-layer 12.0 mm sphere.

The maximum directivity obtained was $D_{\max} = 15.01$ dB with $W = 0.9968$. This value is 7.0 dB larger (5 times larger giving an AE = 500%) than that of the uniformly illuminated circular aperture bound $D_A$, the ideal maximum directivity of this spherical dielectric lens antenna. The parameters of this optimized configuration rounded-off from the optimizer solution values with 16 digits to 4 significant digits are listed in Table II.

TABLE II: **Configuration parameters rounded-off from the 16-digit optimizer solution values for the maximum directivity 10-layer 12.0 mm case to 4 significant digits. The source position is $r_s$ = 13.01 mm and ED : MD = 1 : 0.9968.**

| $i$ | $\varepsilon_r$ | $r$ (mm) |
|---|---|---|
| 1 | 13.97 | 0.104 |
| 2 | 1.000 | 1.799 |
| 3 | 8.380 | 5.256 |
| 4 | 10.70 | 5.736 |
| 5 | 22.66 | 6.423 |
| 6 | 34.73 | 7.825 |
| 7 | 10.62 | 9.087 |
| 8 | 8.149 | 9.540 |
| 9 | 49.38 | 11.10 |
| 10 | 34.86 | 12.00 |

The radius of the smallest enclosing sphere for this case is $a = r_{\text{sys}} = (r_{\max} + r_s)/2$, i.e., the center of the sphere is the midpoint between the source distance $r_s$ and the opposite surface point on the sphere of radius $r_{\text{max}}$. Therefore, the Harrington normal directivity bound is $D_{\text{ka}} = 10\log 10[(kr_{\text{sys}})^2 + 2\ kr_{\text{sys}}] = 10.83$ dB and the Kildal-Best heuristic bound is $D_{\text{KB}} = 10\log 10[(kr_{\text{sys}})^2 + 3] = 9.94$ dB. Consequently, $D_{\text{max}}$ surpasses both of these heuristic bounds by 4.18 and 5.07 dB, respectively. The corresponding aperture efficiencies are AE = 261.82 % and 321.37 %. The other figures of merit for this extreme directivity case are FTBR = 16.77 dB, and SLL = 10.07 dB.

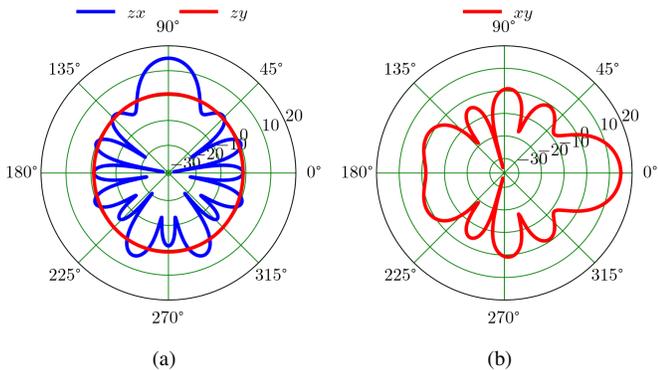

Fig. 5: Directivity patterns of the 10-layer 12.0 mm maximum directivity MMA-spherical lens antenna. (a) Two principal planes (b) Horizontal plane.

While this MMA-spherical lens antenna is superdirective and unidirectional, its directivity patterns, as shown in Fig. 5, have many sidelobes. This is a direct consequence of the higher orders and degrees of the TE and TM modes present within the sphere and, hence, in the overall field exterior to it. In fact, the TM and TE modes involved with these far-field results are illustrated in Fig. 6. The magnitudes of the solution coefficients are plotted relative to their order $n$ and the magnitude of their degree $m$. Since both $\pm m$ degrees always appear together in the VSWEs, we opted to only plot the $|m|$ values to simplify the plot. The magnitudes of each TM and TE mode have been normalized independently relative to that of their largest one. Magnitudes smaller than 0.001 are not plotted which explains why some mode orders appear to be missing.

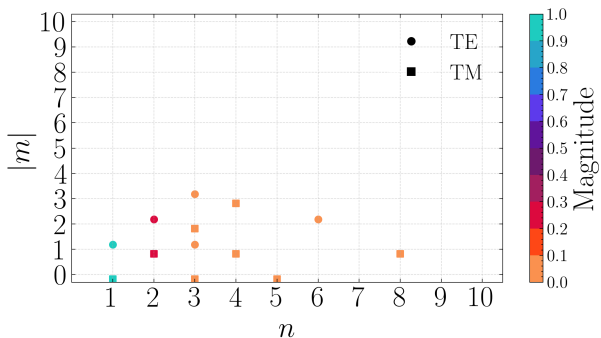

Fig. 6: TM and TE mode decomposition of the optimizer's predicted expansion coefficients in terms of the multipole order and degree and their magnitudes (normalized to the magnitude of the maximum one) for the 10-layer 12.0 mm maximum directivity MMA-spherical lens antenna.

The largest multipole order is $n$ = 8, i.e. the $TM_{81}$ mode. However, if the plotting cutoff is set at 0.05, it is the $TM_{30}$ mode. Thus, we find that the modes of order $n$ = 4 to 8 actually contribute only tiny amounts to the directivity. On the other hand, while the lowest order (dipole) modes have the largest contributions, the multilayers of the sphere clearly supply substantial contributions from the quadrupole and octopole modes as well as at least the presence of yet higher order modes (HOMs). Therefore, the Harrington multipole bound with the practical $N$ = 3 immediately gives $D_{\text{HN}} = N^2 + 2N = 15 = 11.76$ dB. Thus, the optimizer's predicted $D_{\text{max}}$ exceeds not only the $D_A$, $D_{ka}$, and $D_{\text{KB}}$ bounds for its configuration, but also the Harrington multipole bound. Furthermore, the largest order of significance is substantially below $M$ = 15, confirming that this choice was a valid one for our VSWEs. We note that studies are in progress to learn how to establish configurations that specifically emphasize those HOMs since their combinations have the potential to yield yet higher $D_{\text{max}}$ values.

From the experiences associated with the recent development of several superdirective MMAs [41], [63], [86]–[89], it is believed that the Harrington's ultimate bound is exceeded by 3.25 dB simply because there are contributions from other higher order multipoles however small they are. The higher degree contributions for a given order help it reach its Harrington's maximum for each order [90]. Moreover, the tuning of the system's response afforded by the optimized weighting of the ED and MD radiators steps outside of the assumptions made in Harrington's original derivation in which the electric and magnetic multipoles are on equal footing. The actual multipole coefficients are additionally weighted by the independent powers supplied to the ED and MD radiating elements.

Views of the composite mode established within the layers of the sphere are presented in Fig. 7. The magnitude of total field (recall that the total field exterior to the sphere is the combination of the ED, MD, and scattered fields) is plotted in a 20 mm × 20 mm region centered with respect to the origin in each principal coordinate plane. Clearly, this composite mode is resonant.

### D. 3-Layer Extreme FTBR Case

As indicated in Fig. 3, there are numerous superdirective cases with extreme FTBRs. However, the realization of 16 digit design parameters is not feasible from any practical perspective. Consequently, the impact of reducing those optimal values to 4 digits was considered in many cases. Typically the outcomes of the higher number of layer cases, such as the 10-layer extreme directivity case, were more sensitive to this rounding-off of the optimizer-specified design parameters and consequent exterior field solution coefficients. Consequently, we focus here on the 3-layer 12 mm case with the largest FTBR, 118.8 dB, to illustrate that the round-off impact can be negligible from a practical point-of-view. We selected this simpler configuration because it has a very realistic chance of being tested experimentally. This and other 3-layer cases are included as examples that were simulated with HFSS to confirm this insensitivity and will be examined in the next section.

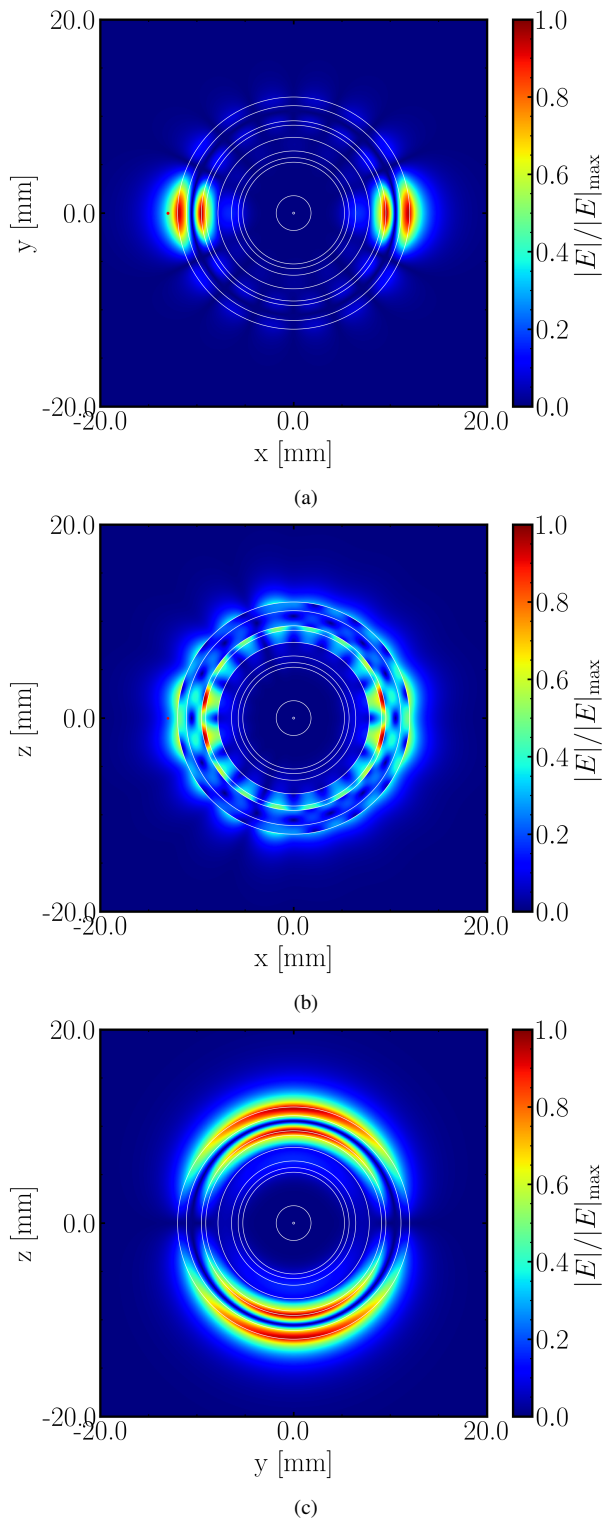

Fig. 7: Normalized magnitude of the total field of the maximum directivity 10-layer 12.0 mm MMA-spherical lens antenna in a 20 mm × 20 mm region centered with respect to the origin in the three principal coordinate planes: (a) $xy$, (b) $xz$, and (c) $yz$. The white circles indicate each layer while the red dot indicates the location of the MMA.

The design parameters for this case rounded-off from the 16-digit optimizer solution values to 4 significant digits are listed in Table III. The MMA is located at $r_\mathrm{s}$ = 14.54 mm and ED:MD = 1.0:1.454, i.e., $W$ = 1.454. The directivity

12TABLE III: **Configuration parameters rounded-off from the 16-digit optimizer solution values for the extreme FTBR 3-layer 12 mm case to 4 significant digits.**

| $i$ | $\varepsilon_\mathrm{r}$ | $r$ (mm) |
|---|---|---|
| 1 | 48.00 | 3.082 |
| 2 | 1.001 | 3.967 |
| 3 | 4.079 | 12.00 |

patterns produced by the 16 digit optimizer design parameters and multipole coefficients are presented in Figs. 8(a) and (b); the corresponding 4 digit ones are presented in Figs. 8(c) and (d).

For the exact optimizer predicted configuration and multipole coefficients $D_\mathrm{max}$ = 12.10 dB, FTBR = 118.8 dB, and SLL = 12.9 dB. Thus, it is superdirective with $D_\mathrm{max} > D_A$ by 4.09 dB and extremely unidirectional. For the rounded-off parameters $D_\mathrm{max}$ = 12.10 dB, FTBR = 81.4 dB, and SLL = 12.9 dB. Thus, the round-off only impacted the FTBR value. Nonetheless, the FTBR remains extreme. It is an excellent candidate for potential experimental verification. Note that while the patterns look identical, please recall that when the directivity is below -35 dB, it is mapped to this value. Thus, the impact of the round-off is truly minimal. Furthermore, examining the normalized magnitude of the VSWE coefficients in the exterior region, the dipole ($n$ = 1) modes remain the most dominant ones, but the quadrupole ($n$ = 2) and octopole ($n$ = 3) modes are found to be actually slightly larger than for the 10-layer extreme directivity case. Note that the Harrington bound $D_\mathrm{HN}$ = 11.76 dB is exceeded by only 0.34 dB. Moreover, since $r_\mathrm{sys}$ = (12+14.54)/2 = 13.27 mm, then $D_{ka}$ = 11.24 dB and $D_\mathrm{KB}$ = 10.31 dB. Thus, $D_\mathrm{max}$ exceeds them both by 0.86 and 1.79 dB, respectively. These results confirm an expected tradeoff – reaching the uppermost maximum directivity comes with a cost of the FTBR.

Views of the composite mode established within the layers of the sphere are presented in Fig. 9 in the same 20 mm × 20 mm region centered with respect to the origin in each principal coordinate plane. Again, this composite mode is quite resonant. Moreover, in contrast to the extreme directivity case, the fields in each of its three layers clearly contribute to the overall behavior of the scattered field response within and near to the lens.

Another interesting aspect of this case is the fact that the ED:MD ratio is ED : MD = 1 : 1.454 while it was ED : MD ≈ 1 : 1 for the maximum directivity case. As will be discussed in regards to the corresponding HFSS simulation results, the MD, in addition to contributing to a large directivity when combined with the ED, has a significant impact on the FTBR. As will be emphasized, the presence of the MD radiated fields and the imbalance between them and the ED radiated ones proves to be very powerful means of tailoring the directivity patterns of the combined MMA-spherical lens antennas.

## V. HFSS Simulations of the Coefficient Resilient Cases

Several of the optimized MMA-spherical lens antennas were considered with practical rather than ideal MMAs and with a limited number of digits for their parameters using ANSYS HFSS (high frequency structure simulator). While HFSS' full-wave finite element (FEM) solutions have proved to be quite accurate in practical engineering design efforts, they are nevertheless numerical variations of ideal cases. They are an important test of how sensitive the optimized configuration parameters are to such variations. Consequently, they provide confirmation that such systems are in fact realizable, i.e., they include the effects of all mutual couplings, material parameters, and impedance matching. Moreover, they also provide engineering guidance as to how the presence of a finite-sized source may affect the spherical lens outcomes in practice. The ED and MD elements of the MMA model in HFSS are driven with lumped sources which themselves are normalized to 1.0 W of input power. Thus a ED:MD ration of 2:1 means the ED and MD elements are excited with 2.0 W and 1.0 W of power, respectively.

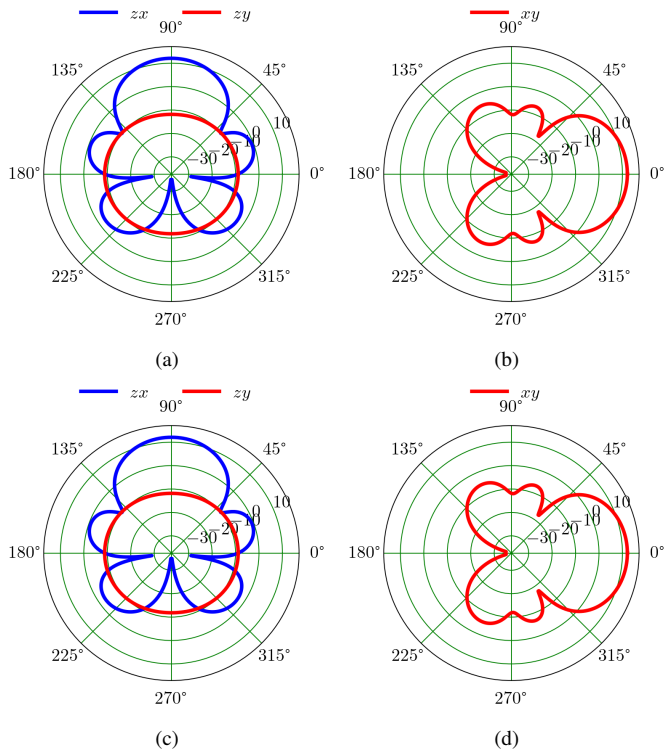

Fig. 8: Directivity patterns of the 3-layer 12 mm extreme FTBR case. Exact 16 digit parameters: (a) $zx$ and $yz$ planes and (b) $xy$ plane. Parameters rounded-off to 4 digits: (a) $zx$ and $yz$ planes and (b) $xy$ plane.

## A. MMA Source Details

The chosen MMA is a generalization of the endfire-radiating ES HDA developed in [91]. As illustrated in Figs. 10(a) and 10(b), it is formed by an ES ED with its $ka_{\mathrm{ED}} = 0.086$ and an ES MD, i.e., a tiny circular loop, with its $ka_{\mathrm{MD}} = 0.063$, which are orthogonal to and in-phase with each other. The length of the dipole along the $z$-axis is 0.825 mm. The loop's outer radius is 0.25 mm. They are separated by a thin rohacell slab with $\varepsilon_r = 1.05$ and dimensions 0.6 mm × 0.825 mm × 0.254 mm. They are both 0.036 mm thick along the $y$-axis away from the rohacell slab, and their traces in the $zx$-plane are 0.025 mm wide. The loop's feedline is 0.10 mm long; the dipole's is 0.3175 mm long. The source gaps along the $z$-axis are 0.05 mm; they are 0.0125 mm wide.

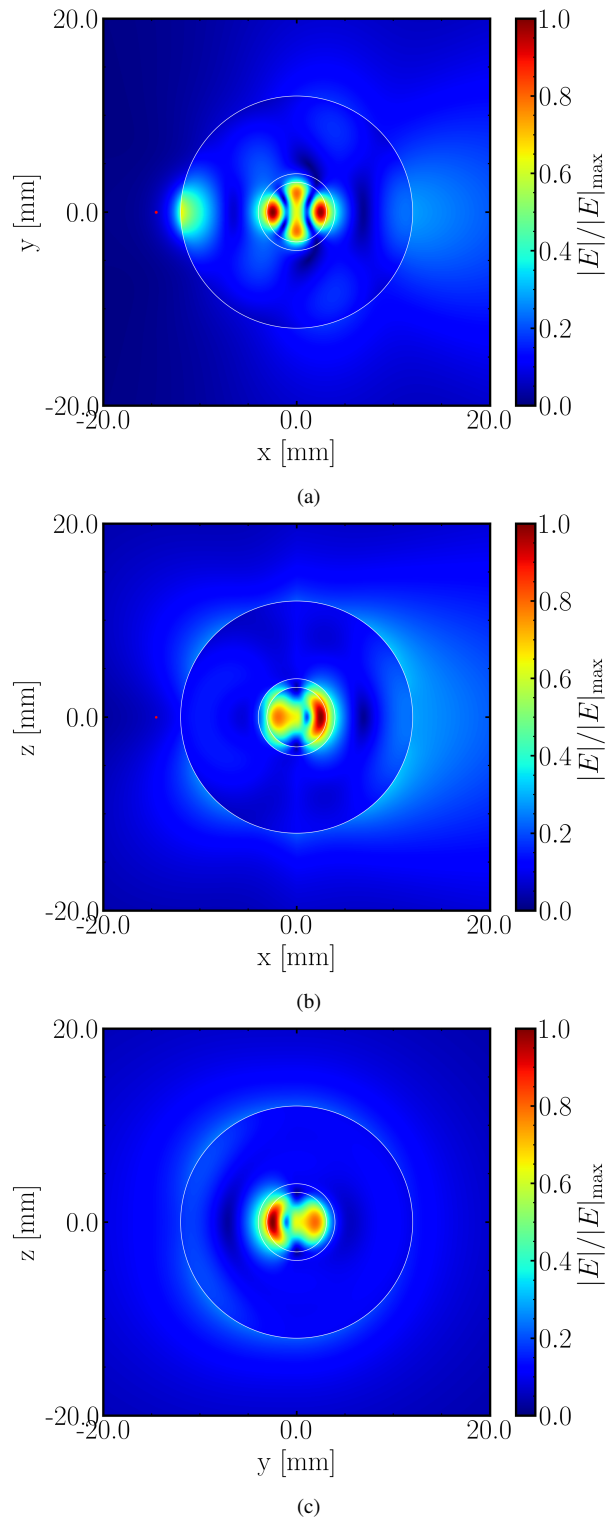

Fig. 9: Normalized magnitude of the scattered field of the 3-layer 12.0 mm extreme FTBR MMA-spherical lens antenna in a 20 mm × 20 mm region centered with respect to the origin in the three principal coordinate planes: (a) $xy$, (b) $xz$, and (c) $yz$. The white circles indicate each layer while the red dot indicates the location of the MMA.

10(c) shows, its maximum directivity, 5.21 dB, is indeed radiated at $\theta = 90°$ in the $\phi = 0°$ plane as designed. Its FTBR was 30.37 dB. Its ED and MD power levels in the presence of the multi-layered sphere were varied manually to optimize the associated performance characteristics.

### B. 3-layer Subwavelength Cases

Because they are the most accessible currently to potential experimental confirmation, the optimized 3-layer 8.5, 10, and 12 mm spheres were simulated. Their optimized design parameters were reduced to the ones indicated in Table IV. Since efficiency was not an issue in this study, all materials were treated as lossless to minimize the simulation times and to simplify the analysis of the results. The table includes the $D_{\max}$ value and the resonance frequency ($f_{\text{res}}$) for which it was obtained; the corresponding FTBR and SLL values; the electrical size $ka_{\text{circ}}$ of their output aperture and the corresponding bound $D_A$; the ED:MD ratio; and, as the principal figure-of-merit (FOM), the difference between $D_{\max}$ and $D_A$ and corresponding AE. The various comparison bounds are also included in which we take into account the entire system where $r_{\max} = r_3$, i.e., $a = r_{\text{sys}} \approx (r_s + r_{\max})/2$ to calculate Harrington's normal directivity and the Kildal-Best bounds.

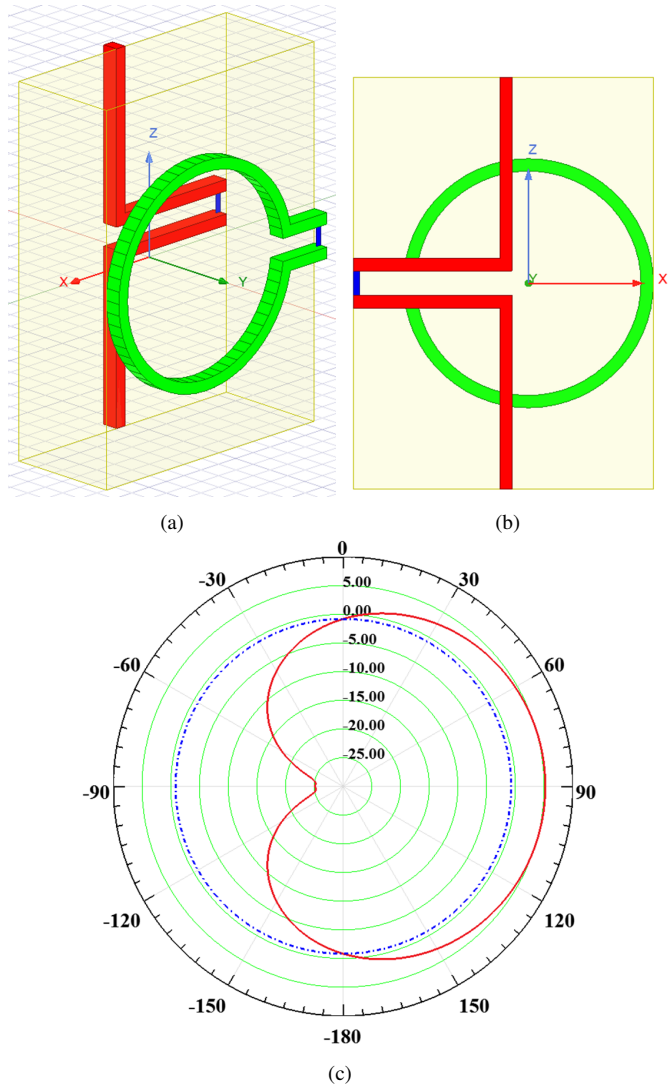

Fig. 10: The endfire radiating MMA source. (a) Dimetric view. (b) Side view ($zx$-plane). (c) Cardioid directivity pattern (dB) in the two principal vertical planes.

TABLE IV: **3-layer Superdirective Spherical Lens Antenna Parameters**

| Parameter | 8.5 | 10 | 12 |
|---|---|---|---|
| $r_1$ (mm) | 1.31 | 2.11 | 3.08 |
| $r_2$ (mm) | 3.69 | 3.19 | 3.97 |
| $a_{\text{circ}} = r_3$ (mm) | 8.5 | 10.0 | 12.0 |
| $\varepsilon_{r1}$ | 92.25 | 52.45 | 48.00 |
| $\varepsilon_{r2}$ | 25.88 | 1.00 | 1.00 |
| $\varepsilon_{r3}$ | 4.29 | 3.85 | 4.08 |
| $r_s$ (mm) | 14.85 | 15.5 | 14.54 |
| $f_{\text{res}}$ (GHz) | 10.10 | 10.17 | 10.09 |
| ED:MD | 1.0:0.8 | 1.0:0.25 | 1.0:2.0 |
| $D_{\max}$ ( dB ) | **10.79** | **11.32** | **12.39** |
| FTBR ( dB ) | 16.87 | 25.20 | 19.91 |
| SLL ( dB ) | 10.54 | 11.90 | 13.53 |
| Aperture electrical size $ka_{\text{circ}}$ | 1.80 | 2.13 | 2.54 |
| $D_A = (ka_{\text{circ}})^2$ ( dB ) | **5.10** | **6.57** | **8.09** |
| FOM ( dB ) | **5.69** | **4.75** | **4.30** |
| AE (%) | 370.50 | 298.29 | 269.23 |
| System electrical size $kr_{\text{sys}}$ | 2.47 | 2.72 | 2.81 |
| $D_{\text{ka}} = (kr_{\text{sys}})^2 + 2\, kr_{\text{sys}}$ (dB) | 10.43 | 11.08 | 11.30 |
| $D_{\text{KB}} = (kr_{\text{sys}})^2 + 3$ (dB) | 9.59 | 10.16 | 10.36 |

The ED and MD are driven independently in the HFSS simulations by lumped sources whose input impedances have been matched to the values predicted by their initial simulations at the frequency of interest. The choice to allow the independent excitation of the ED and MD followed from the discovery that different levels of power radiated by the ED and MD led to enhanced directivity and FTBR levels in presence of the multilayered sphere, i.e., it allowed tailoring the fields radiated by the different stimulated TE and TM modes excited in it as was found to be an asset in the optimized results. The resulting mixture of electric and magnetic multipoles confirms the enhanced superdirective properties that were revealed with the optimized ideal systems. Nevertheless, the MMA was designed as an HDA at 10 GHz, i.e., with 1.0 W delivered to the terminals of both the ED and MD, that radiates maximally along the $+x$-axis. As the cardioid directivity pattern in Fig.

As noted previously, the target frequency for the optimizer was set at 10.0 GHz. The optimizer's 16-digit solution coefficients for a 3-layer $r_{\max}$ = 8.5 mm sphere produced the performance characteristics $D_{\max}$ = 10.60 dB, FTBR = 19.02 dB, and SLL = 10.6 dB with W = 0.88. The associated radii, permittivities, and $r_s$ values that were used in the corresponding HFSS model are listed in Table IV. A side view of the HFSS model is shown in Fig. 11(a). A frequency sweep with W = 1.0 was run to establish the frequency at which the maximum directivity occurred. The input impedances of the MMA' ED and MD elements were then adjusted to achieve matching at that maximum directivity frequency. The consequent resonance at 10.10 GHz is illustrated in Fig. 11(b). The ED:MD ratio impact on the results was then investigated manually. As the ratio was varied, it was found that the frequencies at which the maximum directivity and FTBR values changed insignificantly. The overall best performance based on this process was obtained with $W = 0.8$.

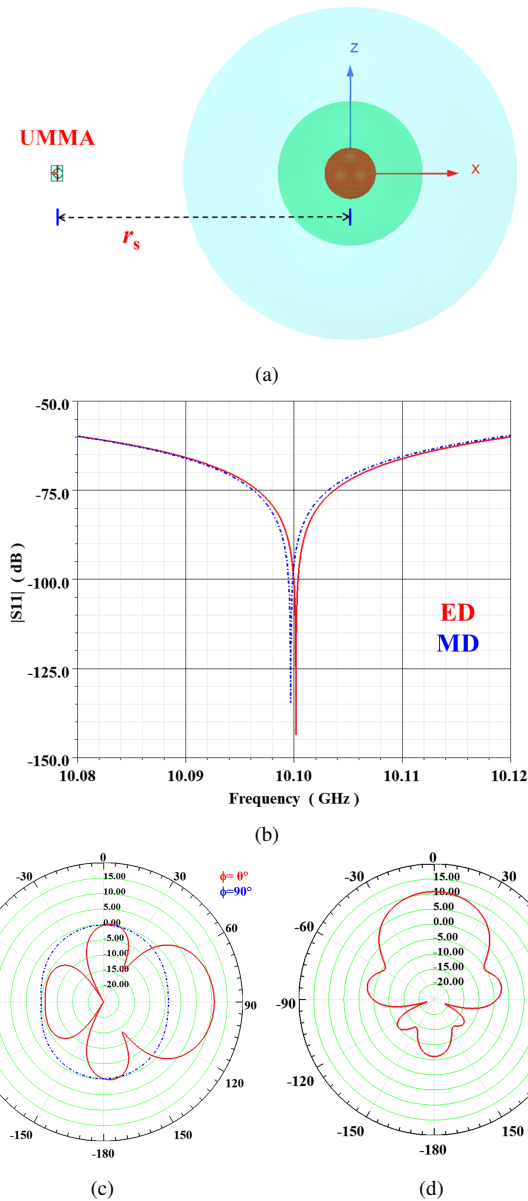

Fig. 11: Directivity patterns of the 8.5 mm 3-layer MMA-spherical lens antenna. (a) HFSS model side view ($zx$-plane). (b) $|S_{11}|$ values as a function of the MMA's frequency. Directivity patterns in (c) the two principal vertical planes and (d) the horizontal plane.

The associated directivity patterns in the two principal vertical planes, $\phi = 0°$ ($zx$-plane) and $\phi = 90°$ ($yz$-plane), and in the horizontal plane, $\theta = 90°$ ($xy$-plane), are presented in Figs. 11(c) and (d), respectively. They were in very reasonable agreement with the optimizer's. The HFSS-predicted figures of merit: $D_{\max} = 10.79$ dB, FTBR = 16.87 dB, and SLL = 10.54 dB. These results are summarized in Table IV. Since $D_{\max}$ exceeds $D_A$ by 5.69 dB, this spherical lens system is clearly superdirective. Moreover, note that it also exceeds both of the heuristic bounds $D_{ka}$ and $D_{\mathrm{KB}}$.

A surprise in the peak directivity outcome is the fact that the rounded-off design parameters and slight change in ED:MD ratio $W$ produced a higher $D_{\max}$ than the optimizer's value. However, note that it came at the cost of lower FTBR and higher SLL values. This result further emphasizes the need to evaluate the optimizer's solution carefully with realistic parameters before assuming the former will immediately happen in practice.

Finally, note that it was found that the HFSS frequency at which $D_{\max}$ was determined was generally slightly greater than the optimizer's target frequency. Furthermore, tuning $W$ led to slightly higher $D_{\max}$ values at the cost of some differences in the FTBR and SLL values. Nonetheless, the cases presented herein all exhibit unidirectional performance with exceptionally large $D_{\max}$ values.

The same procedure was applied to all HFSS cases. The remaining 3-layer cases to discuss herein nicely facilitate demonstrating how the presence of both the ED and MD radiating elements leads to their superdirective outcomes, as well as how their ED:MD ratio impacts their other performance characteristics.

The associated radii, permittivities, and $r_s$ values that were used in the 10 mm spherical lens antenna HFSS model are listed in Table IV. The associated directivity patterns in the two principal vertical planes, $\phi = 0°$ ($zx$-plane) and $\phi = 90°$ ($yz$-plane), and in the horizontal plane, $\theta = 90°$ ($xy$-plane), are presented in Figs. 12(a) and (b), respectively. The HFSS-predicted maximum directivity $D_{\max} = 11.32$ dB, FTBR = 25.20 dB, and SLL = 11.90 dB with $W = 0.25$. These results are summarized in Table IV. While $D_{\max}$ exceeds $D_A$ by 4.75 dB so the system is indeed superdirective, it also exceeds both of the heuristic bounds $D_{ka}$ and $D_{\mathrm{KB}}$.

In comparison, these HFSS results are also in reasonable agreement with the optimizer's results: $D_{\max} = 11.19$ dB and FTBR = 19.18 dB at 10 GHz with $W=0.70$. The directivity patterns are fundamentally quite similar as well.

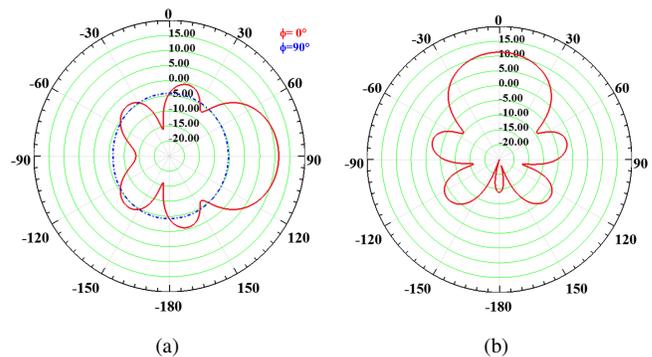

Fig. 12: Directivity patterns of the 10.0 mm 3-layer MMA-spherical lens antenna. (a) Two principal planes (b) Horizontal plane.

The results in Fig. 13 are presented to emphasize the importance of exciting the sphere with both the ED and MD radiated fields. Moreover, they facilitate explaining the impact of the small $W=0.25$ value, i.e., the ED input power was four times that of the MD. The directivity patterns in the two principal vertical planes are shown in Figs. 13(a)-(c) for the ED:MD ratios 1:0, 0:1, and 1:1, respectively. When the MD radiator is turned off, Fig. 13(a) indicates that $D_{\max}$ is reduced by 2.17 dB to 9.15 dB. Moreover, the FTBR reduces to 15.08 dB and the pattern in the $zx$-plane has expanded along the $\pm z$-directions. On the other hand, Fig. 13(b) indicates that when the ED radiator is turned off, the maximum direction of the combined fields is no longer along the $+x$-axis, but rather is back towards the MMA in the $-x$-direction. Moreover, $D_{\max}$ is reduced even more, i.e., by 3.82 dB to 7.50 dB, in that back direction. The associated FTBR = 4.06 dB, and the system is no longer unidirectional. Finally, when the MMA is an HDA, i.e., with $W$ = 1.0, Fig. 13(c) shows that the maximum directivity increased significantly by 2.10 dB from the ED-only excitation to $D_{\max}$ = 11.25 dB with the FTBR = 14.09 dB. This HDA-based system already surpasses all the bounds. It is then clear that emphasizing the ED fields further with $W$ = 0.25, the front lobe is enhanced more and the backlobe is reduced more to yield the final higher $D_{\max}$ and FTBR values. However it is further noted that when $W$ decreases even further, both of these performance characteristics degrade rapidly.

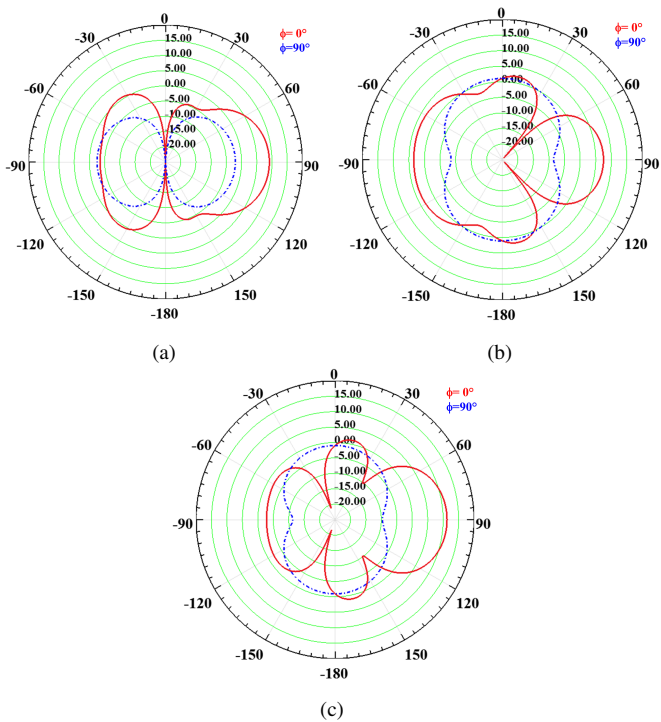

(a) (b)

(c)

Fig. 13: Directivity patterns of the 3-layer 10.0 mm 3-layer MMA-spherical lens antenna for different ED:MD ratios. (a) ED only, i.e., ED:MD = 1:0. (b) MD only, i.e., ED:MD = 0:1. (c) ED and MD with equal power, i.e., ED:MD = 1:1.

The associated radii, permittivities, and $r_s$ values that were used in the 12 mm spherical lens antenna HFSS model are listed in Table IV. This case was selected for HFSS testing because the optimizer's superdirective results included a huge FTBR. In particular, the optimizer predicted $D_{\max}$ = 12.10 dB, FTBR = 120.60 dB, and SLL = 12.85 dB with $W$ = 1.454. The HFSS simulated directivity patterns in the two principal vertical planes, $\phi = 0°$ ($zx$-plane) and $\phi = 90°$ ($yz$-plane), and in the horizontal plane, $\theta = 90°$ ($xy$-plane), are presented in Figs. 14(a) and (b), respectively. The HFSS-predicted performance characteristics were $D_{\max}$ = 12.39 dB, FTBR = 19.91 dB, and SLL = 13.53 dB with $W$ = 2.0. These results are also summarized in Table IV. In this case $D_{\max}$ exceeds $D_A$ by 4.30 dB; the system is indeed superdirective. Furthermore, it also exceeds both of the heuristic bounds $D_{ka}$ and $D_{\mathrm{KB}}$.

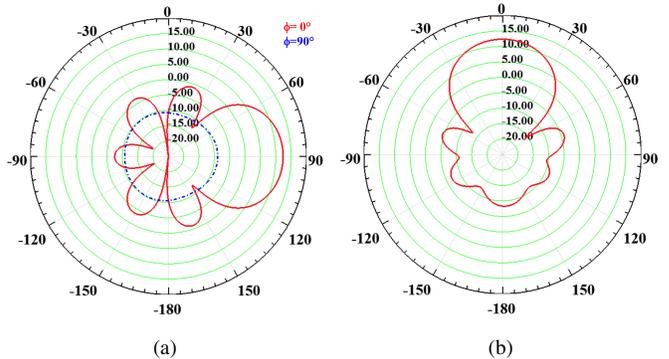

(a) (b)

Fig. 14: Directivity patterns of the 12.0 mm 3-layer MMA-spherical lens antenna. (a) Two principal planes. (b) Horizontal plane.

In comparison to the optimizer's result, $D_{\max}$ exceeds that value by 0.29 dB. However, its FTBR value is substantially smaller but, nonetheless, still confirms that the system is unidirectional. What is noticeably different with the HFSS patterns is that there is a backlobe while the optimizer patterns have quite deep nulls in the back direction. We attribute this feature to the presence of the physically large MMA in the back direction in comparison to the ideal one.

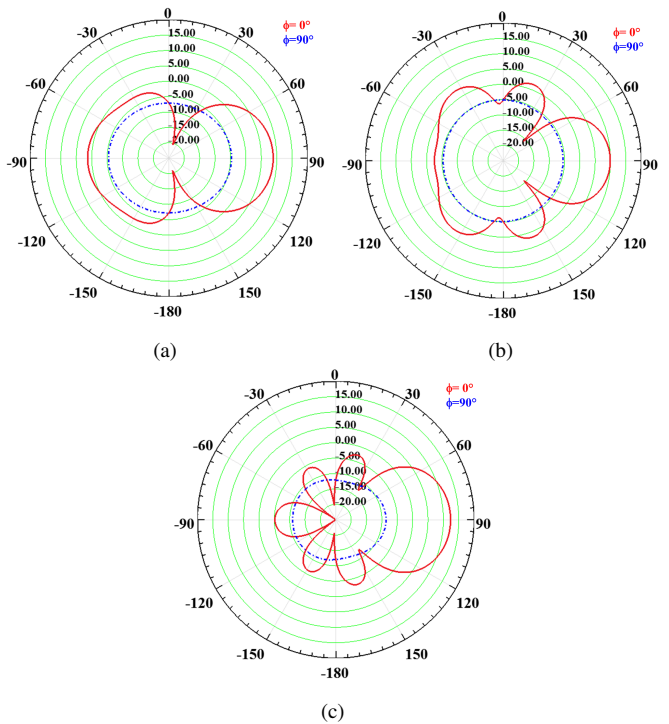

(a) (b)

(c)

Fig. 15: Directivity patterns of the 3-layer 12.0 mm 3-layer MMA-spherical lens antenna for different ED:MD ratios. (a) ED only, i.e., ED:MD = 1:0. (b) MD only, i.e., ED:MD = 0:1. (c) ED and MD with equal power, i.e., ED:MD = 1:1.

The results in Fig. 15 are again presented to emphasize the importance of exciting the sphere with both the ED and MD radiated fields. On the other hand, they facilitate explaining the impact of the now large $W$=2.0 value, i.e., the MD input power was twice that of the ED. The directivity patterns in the two principal vertical planes are shown in Figs. 15(a)-(c) for the ED:MD ratios 1:0, 0:1, and 1:1, respectively. When the MD radiator is turned off, Fig. 15(a) indicates that $D_{\max}$ is reduced by 3.55 dB to 8.84 dB. Moreover, the FTBR reduces significantly to only 7.52 dB and the system is not unidirectional. This occurs because the pattern in the $zx$-plane has expanded in the back -$z$-direction. Figure 15(b) indicates that when the ED radiator is turned off, the maximum direction of the combined fields remains along the +$x$-axis, but has a higher $D_{\max}$ = 9.92 dB that is still 2.47 smaller than the final maximum of the system. On the other hand, this version is again unidirectional with the FTBR = 14.93 dB. Finally, when the MMA is an HDA, i.e., with $W$ = 1.0, Fig. 15(c) shows that the maximum directivity increases significantly, 2.42 dB, from the MD-only excitation, to $D_{\max}$ = 12.33 dB with the FTBR = 17.53 dB and the SLL = 15.15 dB. This HDA-based system already surpasses all the bounds. It is then clear that emphasizing the MD fields further to $W$ = 2.0, the front lobe is enhanced only slightly while the backlobe is further reduced with only a minor cost in the SLL. It is further noted that when $W$ is increased to $W$ = 3.0, the enhanced MD field increases the FTBR to 20.66 but with the cost that $D_{\max}$ decreases to 12.31 dB.

### C. 5-layer Subwavelength Case

As we have illustrated, having more layers can lead to higher directivities. Spherical Luneburg lens (LL) antennas with multiple dielectric layers have been prototyped. For instance, a six-layer 8-$\lambda_0$ diameter LL excited by a waveguide launched field was tested at 6 GHz in [92]. A four-layer 12.93-$\lambda_0$ diameter LL also excited by a waveguide launched field was tested at 38 GHz in [93]. However, the increase of the dielectric constant of each layer in both instances increased only modestly from the core to the outer layer. As our optimized systems have demonstrated, we would want to have no significant constraint on the sequence nor magnitude of the permittivities within some limit from layer to layer. Consequently, we have also considered optimized 5-layer 12 mm configurations and their corresponding HFSS versions. Since the available computing resources limited the size of the HFSS problems that could be simulated, ones with higher numbers of layers were not possible with the appropriate dense meshing required to capture the physics within very thin, rather high permittivity regions.

The optimizer-specified parameters were the radii (in mm) $r_1$ = 1.05, $r_2$ = 2.12, $r_3$ = 3.37, $r_4$ = 3.94, $r_5$ = 12.00; the relative permittivities $\varepsilon_{r1}$ = 74.98, $\varepsilon_{r2}$ = 45.71, $\varepsilon_{r3}$ = 61.13, $\varepsilon_{r4}$ = 2.268, $\varepsilon_{r1}$ = 10.476; and the MMA location $r_s$ = 13.24 mm. The optimizer's performance results were $D_{\max}$ = 12.8 dB with the FTBR = 18.8 dB and the SLL = 16.9 dB with $W$ = 0.64.

The corresponding HFSS simulated directivity patterns in the two principal vertical planes, $\phi$ = 0° ($zx$-plane) and $\phi$ = 90° ($yz$-plane), and in the horizontal plane, $\theta$ = 90° ($xy$-plane), at the resonance frequency $f_{\text{res}}$ = 10.13 GHz are presented in Figs. 16(a) and (b), respectively. The HFSS-predicted maximum directivity is $D_{\max}$ = 12.76 dB with $W$ = 0.8. The associated FTBR = 16.76 dB and SLL = 17.12 dB. These results are actually in quite good agreement with the optimizer's values with $M$ = 15. Moreover, the shapes of the patterns are nicely matched as well.

Furthermore, with the aperture $ka_{\text{circ}}$ = 2.55, the 5-layer system has $D_{\max}$ greater than $D_A$ = 8.12 by 4.67 dB, an increase over the 3-layer 12 mm result by 0.37 dB. Since its $r_s$ is smaller than that case, the system radius reduces to $a = r_{\text{sys}}$ = 12.62 and thus gives $ka$ = 2.68. Consequently, $D_{ka}$ = 10.98 dB and now $D_{\max}$ is greater than it by 1.81 dB. Thus, the 5-layer system provides immediate benefits to the maximum directivity relative to the bounds. Also recall that the 10-layer extreme case's $D_{\max}$ = 15.01 dB, which is greater than its $D_A$ = 8.01 dB by 7.0 dB. These results clearly demonstrate that more layers facilitate the presence of more higher order modes and, hence, the possibility for higher directivities.

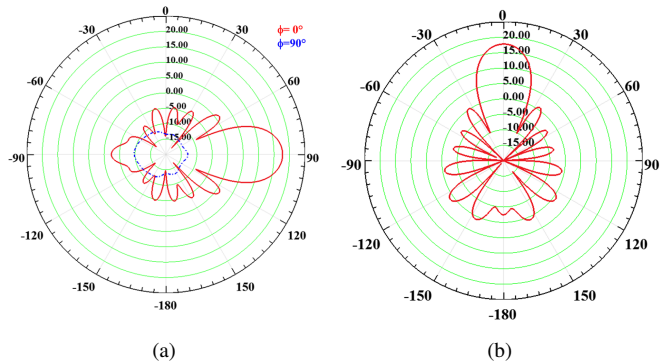

Fig. 16: Directivity patterns of the 12.0 mm 5-layer MMA-spherical lens antenna. (a) Two principal planes (b) Horizontal plane.

We note that we have obtained many other $N$ = 5 cases with higher $D_{\max}$ values. However, they proved to be quite similar and equally sensitive to the number of digits kept for the parameters. For instance, with the reduction of the optimizer's 16 digit results to the rounded-off parameters: radii (in mm) $r_1$ = 0.9932, $r_2$ = 5.9229, $r_3$ = 8.6082, $r_4$ = 11.773, $r_5$ = 12.00; relative permittivities $\varepsilon_{r1}$ = 4.1136, $\varepsilon_{r2}$ = 5.4675, $\varepsilon_{r3}$ = 27.094, $\varepsilon_{r4}$ = 4.4018, $\varepsilon_{r5}$ = 1.4616; MMA location $r_s$ = 12.720 mm; and $W$=0.95154, the performance results were $D_{\max}$ = 13.65 dB with the FTBR = 16.93 dB and the SLL = 10.88 dB, which were quite close to the optimizer's $D_{\max}$ = 13.67 dB with the FTBR = 15.12 dB and the SLL = 10.13. Unfortunately, rounding-off further to 2 or 3 decimal places and even manually trying variations of them, the HFSS results never exceeded $D_{\max}$ = 12.51 dB. Moreover, because of the memory and run-time requirements for these cases, we could only try a few. Basically the optimizer found a local maximum and we could not move it to another less sensitive one that would continue to yield $D_{\max}$ in the mid 13's when evaluated in HFSS. Nevertheless, these additional cases further confirmed the fact that more layers give access to HOMs that can yield higher $D_{\max}$ values.

*D. Wavelength-sized Case*

An ideal $\lambda_0$ radius 3-layer MMA-spherical lens antenna was also considered as a relative comparison to typical current Luneburg lens systems mentioned previously which are many wavelengths in diameter to achieve their high directivities. The configuration of this system, whose radius is much larger that the previous 3-layer cases, employed the optimizer-specified parameters: the radii (in mm) $r_1 = 1.55$, $r_2 = 18.58$, $r_3 = 29.98$; the relative permittivities $\varepsilon_{r1} = 65.24$, $\varepsilon_{r2} = 2.45$, $\varepsilon_{r3} = 2.17$; and the MMA location $r_s = 30.50$ mm. The optimizer's results with $M = 15$ were $D_{\max} = 17.74$ dB and FTBR = 27.93 dB with $W = 0.60$.

The HFSS simulated directivity patterns in the two principal vertical planes, $\phi = 0°$ ($zx$-plane) and $\phi = 90°$ ($yz$-plane), and in the horizontal plane, $\theta = 90°$ ($xy$-plane), are presented in Figs. 17(a) and (b), respectively. The HFSS-predicted maximum directivity $D_{\max} = 17.78$ dB occurs at 10.37 GHz with $W = 0.40$. The FTBR = 20.25 dB and SLL = 16.57 dB. With $ka_{\text{circ}} = 6.52$, the bound $D_A = 16.28$ dB. Thus, $D_{\max}$ exceeds $D_A$ by 1.50 dB, and the system is superdirective. With $r_{\text{sys}} = 30.24$, $ka = 6.57$ and thus $D_{ka} = 17.51$ dB and $D_{\text{KB}} = 16.65$ dB. Consequently, $D_{\max}$ also exceeds both of these heuristic bounds by 0.27 and 1.13 dB, respectively. Moreover, the AE = 141 %.

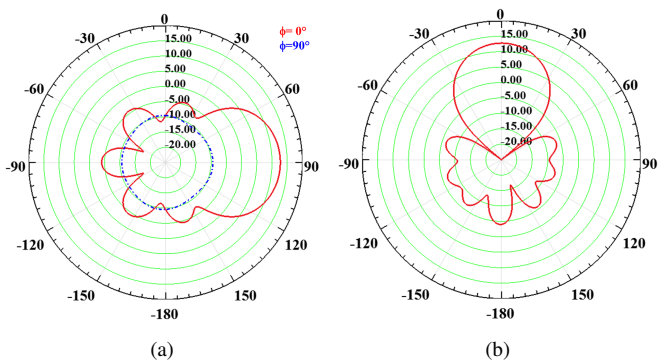

Fig. 17: Directivity patterns of the $\sim \lambda_0$ radius ($\sim 30$ mm) 3-layer MMA-spherical lens antenna. (a) Two principal planes (b) Horizontal plane.

Finally, a corresponding $\lambda_0$ radius 8-layer MMA-spherical lens antenna was completed. It took several days of run time to finish a single configuration with a discrete sweep including just a few frequency points. The problem size max'd out the available physical memory and the code ran with virtual paging. The frequency region in which the maximum occurred was estimated properly based on the 3-region version. The HFSS results were nevertheless in reasonable agreement with the optimizer results.

The rounded-off parameters of the optimizer's 16 digit results were: radii (in mm) $r_1 = 0.775$, $r_2 = 8.648$, $r_3 = 11.946$, $r_4 = 15.901$, $r_5 = 17.672$, $r_6 = 25.367$, $r_7 = 26.31$, $r_8 = 29.979$; relative permittivities $\varepsilon_{r1} = 1.23$, $\varepsilon_{r2} = 8.572$, $\varepsilon_{r3} = 9.477$, $\varepsilon_{r4} = 15.006$, $\varepsilon_{r5} = 34.244$, $r_6 = 33.272$, $r_7 = 49.999$, $r_8 = 5.587$; MMA location $r_s = 29.9959$ mm; and $W = 0.9405$. The optimizer-predicted performance figures-of-merit were $D_{\max} = 18.91$ dB with the FTBR = 23.97 dB at 10.0 GHz.

Unfortunately, the optimizer specified distance $r_s$ was much too close to the outer surface of the spherical lens to allow for the presence of the finite-sized MMA in the HFSS model. It was determined that the small alternate MMA location with $r_s = 30.3$ mm was acceptable to the software. The simulation then proceeded. With rather time consuming manual tuning of the ED:MD ratio, the best value was determined to be 1.4:1, i.e., $W = 0.7143$.

The maximum directivity $D_{\max} = 18.43$ dB was obtained at 10.13 GHz. The corresponding FTBR = 23.05 dB with the SLL = 20.07 dB in the forward hemisphere and 17.85 in the back one. The HFSS simulated directivity patterns in the two principal vertical planes, $\phi = 0°$ ($zx$-plane) and $\phi = 90°$ ($yz$-plane), and in the horizontal plane, $\theta = 90°$ ($xy$-plane), are presented in Figs. 18(a) and (b), respectively. Thus, $D_{\max}$ exceeds $D_A = 16.08$ by 2.35 dB, and the system is superdirective. With $r_{\text{sys}} = 30.15$, $ka = 6.40$ and thus $D_{ka} = 16.13$ dB and $D_{\text{KB}} = 16.43$ dB. Moreover, the AE = 172%. Consequently, $D_{\max}$ again exceeds both of these heuristic bounds, but now by larger amounts, i.e., 2.3 and 2.0 dB, respectively. These outcomes further illustrate that with more properly designed layers, higher directivities are attainable. We expect with more higher-performance computing (HPC) resources that we could more thoroughly examine these $\lambda$-radius systems as well as explore the performance of yet even larger ones comparable in size to the many current multiple wavelength in diameter Luneburg lenses.

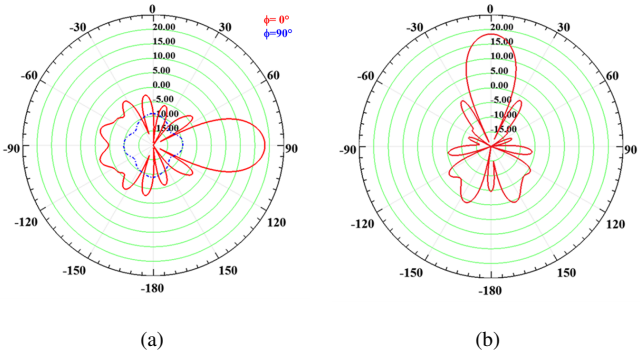

Fig. 18: Directivity patterns of the $\sim \lambda_0$ radius ($\sim 30$ mm) 8-layer MMA-spherical lens antenna. (a) Two principal planes (b) Horizontal plane.

## VI. CONCLUSIONS

Several superdirective unidirectional MMA-spherical lens antennas were reported. Basic MMAs formed by a combination of independently driven ED and MD elements excited different sized multi-layered dielectric spheres composed of low and high dielectric constant layers. The configurations in which the MMA consisted of ideal ED and MD elements were optimized with a genetic algorithm (GA) approach whose cost-functions emphasized the maximum directivity and FTBR values. Several cases were then confirmed with HFSS simulations in which a practical, finite-sized MMA excited the spheres. Their superdirective properties were confirmed as well as the fact that they also achieved $D_{\max}$ values that exceeded the various popular directivity bounds found in the literature. It was demonstrated why the presence of both the ED and MD radiating elements and their relative weighting facilitated their exceptional performance characteristics. A fundamental reason why this occurs is the fact that Harrington's bounds assume that the electric and magnetic multipoles are excited with the same weights. We have broken this assumption by allowing them to be excited with different weights.

Preliminary realizable configurations of some of the many cases reported herein are currently being considered for their experimental verification. This includes accounting for the dielectric losses. They were ignored in this study given the physics and optics literature acknowledging them to be generally negligible, particularly at microwave frequencies. Because loss values are highly dependent on the specific materials employed for a practical realization, potential prototypes are being explored carefully with HFSS simulations to meet standard engineering fabrication and measurement criteria. Initial simulations indicate that they will have an impact primarily on the frequencies at which the maximum figures-of-merit are attained. Simultaneously, the possibility of using graphics processing unit (GPU) accelerators are also being considered. They are commonly acknowledged to be an affordable means to speed up parallel computations [94]. It is anticipated that practical superdirective MMA-spherical dielectric lens antennas based on the concepts and examples described in this paper could readily meet the growing demands for the highly directive systems envisioned for many NextG wireless communications, sensor networks, and battery-less (wireless power transfer) applications.

ACKNOWLEDGMENTS

The authors gratefully acknowledge the computational and data resources provided on the Sophia HPC Cluster at the Technical University of Denmark, DOI:10.57940/FAFC-6M81.